\documentclass[preprint,showpacs,superscriptaddress,longbibliography]{revtex4-1}
\usepackage{diagbox}
\usepackage{amsmath}
\usepackage{amsfonts}
\usepackage{physics}
\usepackage[dvipsnames]{xcolor}
\usepackage[space]{grffile}
\usepackage[hidelinks]{hyperref}
\usepackage{natbib}
\usepackage{braket}
\usepackage{float}
\usepackage{color}
\usepackage{blindtext}
\usepackage{comment}

\begin{document}

\title{A Hubbard exciton fluid in a photo-doped antiferromagnetic Mott insulator}

\author{Omar Mehio}
\thanks{These authors contributed equally to this work.}
\affiliation{Institute for Quantum Information and Matter, California Institute of Technology, Pasadena, CA 91125}
\affiliation{Department of Physics, California Institute of Technology, Pasadena, CA 91125}

\author{Xinwei Li}
\thanks{These authors contributed equally to this work.}
\affiliation{Institute for Quantum Information and Matter, California Institute of Technology, Pasadena, CA 91125}
\affiliation{Department of Physics, California Institute of Technology, Pasadena, CA 91125}

\author{Honglie Ning}
\thanks{These authors contributed equally to this work.}
\affiliation{Institute for Quantum Information and Matter, California Institute of Technology, Pasadena, CA 91125}
\affiliation{Department of Physics, California Institute of Technology, Pasadena, CA 91125}

\author{Zala Lenarčič}
\affiliation{Department for Theoretical Physics, Jozef Stefan Institute, Ljubljana, Slovenia}
\affiliation{Department of Physics, University of California, Berkeley, CA 94720, USA}

\author{Yuchen Han}
\affiliation{Institute for Quantum Information and Matter, California Institute of Technology, Pasadena, CA 91125}
\affiliation{Department of Physics, California Institute of Technology, Pasadena, CA 91125}

\author{Michael Buchhold}
\affiliation{Institute for Theoretical Physics, University of Cologne, Cologne, Germany}

\author{Zach Porter}
\affiliation{Materials Department, University of California, Santa Barbara, CA 93106}

\author{Nicholas J. Laurita}
\affiliation{Institute for Quantum Information and Matter, California Institute of Technology, Pasadena, CA 91125}
\affiliation{Department of Physics, California Institute of Technology, Pasadena, CA 91125}

\author{Stephen D. Wilson}
\affiliation{Materials Department, University of California, Santa Barbara, CA 93106}

\author{David Hsieh}
\email[Author to whom the correspondence should be addressed: ]{dhsieh@caltech.edu}
\affiliation{Institute for Quantum Information and Matter, California Institute of Technology, Pasadena, CA 91125}
\affiliation{Department of Physics, California Institute of Technology, Pasadena, CA 91125}

\maketitle

\noindent\textbf{Abstract}

\textbf{The undoped antiferromagnetic Mott insulator naturally has one charge carrier per lattice site. When it is doped with additional carriers, they are unstable to spin fluctuation-mediated Cooper pairing as well as other unconventional types of charge, spin, and orbital current ordering. Photo-excitation can produce charge carriers in the form of empty (holons) and doubly occupied (doublons) sites that may also exhibit charge instabilities. There is evidence that antiferromagnetic correlations enhance attractive interactions between holons and doublons, which can then form bound pairs known as Hubbard excitons, and that these might self-organize into an insulating Hubbard exciton fluid. However, this out-of-equilibrium phenomenon has not been detected experimentally. Here, we report the transient formation of a Hubbard exciton fluid in the antiferromagnetic Mott insulator Sr$_{2}$IrO$_{4}$ using ultrafast terahertz conductivity. Following photo-excitation, we observe rapid spectral weight transfer from a Drude metallic response to an insulating response. The latter is characterized by a finite energy peak originating from intra-excitonic transitions, whose assignment is corroborated by our numerical simulations of an extended Hubbard model. The lifetime of the peak is short, approximately one picosecond, and scales exponentially with Mott gap size, implying extremely strong coupling to magnon modes.} \\ 

\noindent\textbf{Main Text}

In weakly correlated rigid band insulators, the Coulomb attraction between photo-doped conduction band electrons and valence band holes leads to the formation of bound excitons. Depending on the structural and electronic properties of the host material, a wide array \cite{regan_emerging_2022, tartakovskii_excitons_2020, dimitriev_dynamics_2022, bae_anisotropic_2022, baranowski_excitons_2020, miller_excitons_1985, bardeen_structure_2014} of exciton species characterized by different spatial distributions and internal energy spectra (both adhering to \cite{kazimierczuk_giant_2014, jadczak_exciton_2019} and strongly departing from \cite{chernikov_exciton_2014, ye_probing_2014} a hydrogen-like progression) can be realized. This phenomenological diversity, coupled with the ability to engineer and externally perturb materials, has created a vast range of optical and opto-electronic functionalities and corresponding device applications \cite{mak_photonics_2016}. Moreover, ensembles of excitons can be prepared through photo-excitation or collective electronic instabilities, creating a platform to study quantum many-body phenomena, including Bose-Einstein condensation \cite{snoke_spontaneous_2002}, insulator-to-metal transitions \cite{kaindl_ultrafast_2003, zhang_stability_2016}, density wave ordering \cite{Butov2002} and electron-hole droplet formation \cite{a_la_guillaume_electron-hole_1983}.

Excitons in strongly correlated electron systems are a subject of growing interest because complex interactions between charge, orbital, spin and lattice degrees of freedom in such materials can drive excitonic effects that do not exist in rigid band insulators. For example, coupling between Coulomb-bound excitons and magnetic degrees of freedom in the host crystal can lead to magnon-exciton bound states \cite{gnatchenko_exciton-magnon_2011, meltzer_exciton-magnon_1968}, exciton-based coherent magnon sensing \cite{bae_exciton-coupled_2022} and the formation of coherent many-body excitonic states \cite{kang_coherent_2020}. In the particular case of two-dimensional (2D) antiferromagnetic (AFM) Mott insulators, the strong interplay between photo-doped carriers and the underlying AFM order can lead to an exciton binding mechanism that is fundamentally different from standard Coulomb attraction. Spatially separating an empty (holon) and doubly occupied (doublon) site - created via photo-excitation resonant with the Mott gap - leaves a string of defects in the AFM background that induces a confining potential \cite{lenarcic_charge_2014, lenarcic_ultrafast_2013, huang_spin-mediated_2020, shinjo_density-matrix_2021, wrobel_excitons_2002,  clarke_particle-hole_1993, grusdt_pairing_2022}. A holon-doublon (HD) pair that is bound through such a magnetic exchange mediated attractive interaction is dubbed a Hubbard exciton (HE) \cite{clarke_particle-hole_1993, grusdt_pairing_2022}. Since the properties of HEs, such as their binding energies, effective masses and lifetimes, are predicted to depend sensitively on the AFM correlations \cite{clarke_particle-hole_1993, tohyama_exact_2005, huang_spin-mediated_2020, shinjo_density-matrix_2021, lenarcic_charge_2014, lenarcic_ultrafast_2013, grusdt_pairing_2022}, 2D Mott AFMs 
may be a platform for realizing magnetically tunable excitonic
devices and collective excitonic phenomena that are inaccessible using the Coulomb-bound excitons found in rigid band insulators or spin-charge separated systems such as one-dimensional Mott
insulators \cite{Penc_96, Jeckelmann_2001, wall_quantum_2011, miyamoto_biexciton_2019}.

HEs have been extensively studied theoretically \cite{clarke_particle-hole_1993, tohyama_exact_2005, huang_spin-mediated_2020, shinjo_density-matrix_2021, lenarcic_charge_2014, lenarcic_ultrafast_2013, grusdt_pairing_2022}, with intriguing parallels to the problem of Cooper-pair formation in Mott AFMs \cite{grusdt_pairing_2022}. Experimentally, however, many basic questions about their properties remain unresolved, including their energy spectra and wavefunctions. Critically, their existence as stable quasiparticles is an open question. Unlike conventional excitons in rigid band insulators or Coulomb-bound excitons in correlated insulators, which appear in interband optical spectra as sharp peaks that are well-separated from the free-particle continuum \cite{MILLER1985520, kazimierczuk_giant_2014, ye_probing_2014, kang_coherent_2020, bae_exciton-coupled_2022, gnatchenko_exciton-magnon_2011, meltzer_exciton-magnon_1968, miyamoto_biexciton_2019}, potential HE resonances in AFM Mott insulators \cite{terashige_doublon-holon_2019, gretarsson_crystal-field_2013, wang_momentum-dependent_1996, gossling_mott-hubbard_2008, alpichshev_confinement-deconfinement_2015, novelli_ultrafast_2012, lovinger_influence_2020} have often been reported to manifest as broad peaks that overlap with the Mott gap edge \cite{shinjo_density-matrix_2021, gretarsson_crystal-field_2013, wang_momentum-dependent_1996, gossling_mott-hubbard_2008}. This lack of separation implies that the excitons are unstable against decay into the free HD continuum and that a bound state is possibly not well defined. Therefore, whether a HE can exist in 2D AFM Mott insulators as a bound state remains unclear.

A direct approach to distinguishing between an unbound HD plasma and a bound HE fluid is to exploit their unique low energy spectral features at THz frequencies. Whereas the former exhibits a metallic Drude response, the latter should exhibit an insulating response characterized by finite energy peaks corresponding to transitions between different excitonic bound states (Figure 1a), which are predicted to lie in the meV range \cite{kaindl_ultrafast_2003, zhang_stability_2016, wrobel_excitons_2002, clarke_particle-hole_1993}. Tracking HE dynamics via intra-excitonic peaks, rather than via excitonic resonances in interband optical spectra, is also advantageous because it enables access to optically dark HEs with finite center-of-mass momenta and is not obscured by the effects of photo-doping induced bandgap renormalization and mid-gap state formation, which are pronounced in Mott insulators \cite{okamoto_photoinduced_2011}. 

Here, we use time-resolved time-domain THz spectroscopy (tr-TDTS) to probe the transient low energy dynamics of photo-doped holons and doublons in the square lattice AFM Mott insulator Sr$_{2}$IrO$_{4}$. The low energy electronic structure of Sr$_{2}$IrO$_{4}$ consists of a completely filled band of spin-orbital entangled pseudospin $J_{eff} = 3/2$ states and a narrow half-filled band of $J_{eff} = 1/2$ states, which splits into a lower Hubbard band (LHB) and an upper Hubbard band (UHB) due to on-site Coulomb repulsion \cite{kim_novel_2008}. Optical conductivity measurements show the LHB $\rightarrow$ UHB transition peak (dubbed the $\alpha$ transition) lying just below 0.6 eV (Figure 1b, c) \cite{moon_temperature_2009, seo_infrared_2017}. The localized $J_{eff} = 1/2$ moments are coupled through strong Heisenberg-type exchange interactions (\textit{J} = 60 meV) \cite{kim_magnetic_2012}, which is in principle conducive to spin-mediated HD binding, and undergo long-range Néel-type AFM ordering below a temperature $T_{N}$ = 230 K \cite{cao_weak_1998}. Although resonant inelastic x-ray scattering measurements have shown evidence of a spin-orbital resonance inside the HD continuum, associated with an intra-site $J_{eff} = 3/2$ to $J_{eff} = 1/2$ excitation \cite{kim_magnetic_2012},  no direct evidence of HEs - generated through inter-site LHB to UHB excitation - has been reported. 

Ultrafast tr-TDTS measurements were performed in reflection geometry on (001) single crystals of Sr$_{2}$IrO$_{4}$ (Methods). HD pairs are optically generated using a near-infrared 100 fs (FWHM) pump pulse tuned on-resonance with the $\alpha$ transition (0.6 eV). After a variable time delay $t$, the low energy charge response is probed by a phase-locked broadband (2 $\rightarrow$ 24 meV) THz pulse, whose electric field profile in the time domain is measured by electro-optic sampling (EOS) as a function of the recording time $t_{\textrm{EOS}}$ (Figure 1d). Figure 1e shows the reflected THz field transient \textit{E}($t_{\textrm{EOS}}$) from the un-pumped crystal overlaid with its pump-induced change $\Delta$\textit{E}($t_{\textrm{EOS}}$, $t$) recorded at a fixed time delay of $t$ = 2.55 ps. The predominant features of $\Delta$\textit{E}($t_{\textrm{EOS}}$, $t$) track \textit{E}($t_{\textrm{EOS}}$) with minimal phase offset, indicating that the presence of photo-dopants increases the THz reflectance as expected. 

Pump-induced differential THz spectra in the frequency domain are obtained by Fourier transforming the field transients with respect to $t_{\textrm{EOS}}$. Figure 1f shows typical spectra from Sr$_{2}$IrO$_{4}$ acquired at a temperature $T$ = 80 K and a pump fluence of 2 mJ/cm$^{2}$ plotted as a function of $t$. At all frequencies, we observe a fast rise in the reflected field amplitude upon injection of photo-dopants at $t$ = 0, followed by a slower exponential decay over several picoseconds. Four peaks clearly appear in the spectra after pump-excitation. There are three sharp peaks above 3 THz that match the three lowest frequency infrared-active phonon modes in Sr$_{2}$IrO$_{4}$ \cite{moon_temperature_2009, seo_infrared_2017, dashwood_momentum-resolved_2019}, as well as one broader peak centered near 1.5 THz, which lies more than 1 THz above the highest energy zone-center magnon mode \cite{kim_magnetic_2012, dean_ultrafast_2016, bahr_low-energy_2014}. No other finite energy peaks were observed at lower frequencies down to 0.35 THz (Supplementary Information Section I). Our measurement of the refractive index of un-pumped Sr$_{2}$IrO$_{4}$ by TDTS in transmission geometry (Methods) shows no evidence of phonon or magnon peaks in the vicinity of 1.5 THz (Figure 1c), consistent with previous reports. These results suggest that the broad peak is not an excitation of a structural or magnetic origin, and can potentially be attributed to an excitonic origin (Supplementary Information Section II). 

This conclusion is supported through an analysis of the full dielectric response of the photo-doped state, which is extracted from the differential THz spectra through standard electrodynamical relations using the thin film approximation (Methods, Supplementary Information Section III). Figures 2a-f show the pump-induced change to the real $\Delta\sigma_{1}(\omega)$ and imaginary $\Delta\sigma_{2}(\omega)$ parts of the optical conductivity at different $t$. Immediately upon injection of photo-dopants ($t$ = 0) there is a positive increase in both $\Delta\sigma_{1}(\omega)$ and $\Delta\sigma_{2}(\omega)$, as is expected from a conducting plasma of unbound holons and doublons \cite{okamoto_photoinduced_2011, zhang_stability_2016, steinleitner_direct_2017}. Over the next several hundred femtoseconds, $\Delta\sigma_{1}(\omega)$ evolves into a peak shape centered around 1.5 THz, while $\Delta\sigma_{2}(\omega)$ develops a dispersive lineshape with a zero-crossing at the same frequency. These are signatures of a Lorentzian dielectric function, consistent with the absorptive response of a bound Hubbard excitonic state. In fact, this evolution from a Drude-like response into an absorptive Lorentzian response is a hallmark of exciton formation following photo-excitation resonant with the $\alpha$ transition, with the Lorentzian representing a transition from one internal excitonic level to another \cite{kaindl_ultrafast_2003, zhang_stability_2016, steinleitner_direct_2017}. The Lorentzian response is diminished relative to the Drude peak when the pump photon energy is tuned to the $J_{eff}$ = 3/2 $\rightarrow$ UHB transition (Supplementary Information Section IV), further supporting its excitonic origin.

To understand the temporal interplay between free and bound HD states, $\Delta\sigma_{1}(\omega)$ and $\Delta\sigma_{2}(\omega)$ were simultaneously fit to a sum of Drude and Lorentz oscillator functions at each $t$ (Methods, Extended Data Figure 1). The spectral weights (SW) associated with the Drude and intra-excitonic Lorentzian components, defined as the area under the fits to $\Delta\sigma_{1}(\omega)$ (Methods), are proportional to the number of free and bound carriers respectively \cite{kaindl_transient_2009}. As shown in Figure 2g, the Drude SW increases from zero starting at $t$ = 0 and reaches a maximum ($t \approx$ 0.4 ps) after the pump pulse has been completely absorbed. Previous time-resolved near-infrared reflectivity \cite{hsieh_observation_2012, dean_ultrafast_2016} and angle-resolved photoemission spectroscopy measurements \cite{piovera_time-resolved_2016} on Sr$_{2}$IrO$_{4}$ showed that intraband cooling of photo-dopants occurs on an ultrashort timescale of around 60 fs, which is typical for AFM Mott insulators \cite{okamoto_photoinduced_2011}. Therefore, by $t$ = 0.4 ps, the unbound holons and doublons have relaxed near the Hubbard band edges. At these early times, the system exhibits a finite Drude SW with a relatively low scattering rate (3.9 $\pm$ 0.1 THz at $t$ = 0.3 ps) compared to 15 $\%$ Rh doped metallic Sr$_{2}$IrO$_{4}$ crystals \cite{xu_optical_2020}, indicating conducting behavior. Between $t$ = 0.4 ps and 1 ps, there is a rapid decay of the Drude SW that coincides with a rise in the Lorentzian SW from zero, demonstrating a SW transfer from the former to the latter component. This aligns with our expectation that free holons and doublons can only bind into stable HEs when their kinetic energy falls below a threshold value. At $t$ = 1 ps, the Lorentzian SW reaches a maximum and subsequently decays over a timescale of several picoseconds due to HE recombination. During this decay, there is a time window when the system possesses a finite Lorentzian SW but vanishing Drude SW within our experimental resolution, thus realizing a transient insulating HE fluid. A dynamical crossover from a conducting electron-hole plasma to an insulating excitonic fluid was previously identified in the photo-doped rigid band semiconductor GaAs through analogous features in tr-TDTS data \cite{kaindl_ultrafast_2003, zhang_stability_2016}. However, the characteristic timescales for exciton formation and decay in Sr$_{2}$IrO$_{4}$ are three orders of magnitude shorter. 

Since tr-TDTS probes excitons with center-of-mass momenta lying both inside and outside of the light cone \cite{kaindl_ultrafast_2003, poellmann_resonant_2015}, it is sensitive to all radiative and non-radiative recombination pathways. In WSe$_2$, for example, an ultrafast radiative recombination of bright excitons and a slower Auger recombination of dark excitons was clearly manifested through a two-step decay of the 1$s$-2$p$ intra-excitonic peak \cite{poellmann_resonant_2015}. To uncover the pathway underlying the ultrafast HE recombination in Sr$_{2}$IrO$_{4}$, we measured the decay dynamics of the total SW about the 1.5 THz mode - obtained by integrating $\Delta\sigma_{1}(\omega)$ from 0.8 to 2 THz - at a series of temperatures from 80 K to 300 K (Figure 3a). The functional form of the decay for $t >$ 1 ps, which is dominated by the Lorentzian term (Figure 2g), is well described by a single exponential at all temperatures (Figure 3a, Methods), suggesting one predominant exciton recombination pathway. Although sub-picosecond radiative recombination has been reported in semiconductors with large interband dipole moments \cite{poellmann_resonant_2015,steinleitner_direct_2017}, this process is unlikely in Sr$_{2}$IrO$_{4}$ owing to the $d$-$d$ character of the $\alpha$ transition \cite{kim_novel_2008} and the complete absence of excitonic peaks in interband optical spectra \cite{seo_infrared_2017, moon_temperature_2009} (Figure 1b). Auger recombination can also be ruled out because the exponential decay time is fluence independent (Extended Data Figure 2), indicating a monomolecular rather than multi-molecular recombination process.  

Given the strong coupling of charges to the pseudospin \cite{moon_temperature_2009, kim_magnetic_2012} and lattice \cite{moon_temperature_2009, hu_spectroscopic_2019, li_strong_2016} degrees of freedom in Sr$_{2}$IrO$_{4}$, we examine the possible role of collective bosonic excitations in the recombination of HEs (Supplementary Information Section II). The highest energy magnon \cite{kim_magnetic_2012} and phonon \cite{dashwood_momentum-resolved_2019} modes in Sr$_{2}$IrO$_{4}$ lie near 200 meV and 90 meV respectively, well below the Mott gap scale (Figure 1b). Therefore HEs can in principle recombine through multi-magnon or higher-order multi-phonon emission channels \cite{lenarcic_charge_2014, lenarcic_ultrafast_2013, lenarcic_exciton_2015}. This mechanism naturally explains the strong temperature dependence of the exponential HE decay time $\tau$, which clearly contrasts with the largely temperature independent decay of the infrared charge response (Figure 3b, Extended Data Figure 3), for the following reason. Optical conductivity measurements have shown that upon cooling from 300 K to 80 K, the Mott gap $\Delta(T)$ of Sr$_{2}$IrO$_{4}$ is significantly enhanced \cite{moon_temperature_2009, seo_infrared_2017}. This means that a greater number of bosons must be emitted in order for HEs to relax across the Mott gap, leading to a larger $\tau$ \cite{strohmaier_observation_2010}. In fact, numerical studies of the 2D square lattice Hubbard model have predicted that HEs can rapidly recombine through multi-magnon emission on the picosecond timescale with $\tau\propto \exp\left[\zeta\Delta(T)/J\right]$, where $J$ is the magnetic exchange energy and $\zeta$ is a factor of order unity \cite{lenarcic_ultrafast_2013, lenarcic_charge_2014}. Using the experimentally reported values of $\Delta(T)$ \cite{moon_temperature_2009} and $J$ = 60 meV \cite{kim_magnetic_2012}, we find that this equation provides a good fit to the measured temperature dependence of $\tau$ (Figure 3b), with a fitted value of $\zeta = 0.76(2)$.This recombination mechanism may be further corroborated in future by direct detection of the transient bosonic excitation spectra in Sr$_{2}$IrO$_{4}$ using techniques such as time-resolved resonant inelastic x-ray \cite{dean_ultrafast_2016} or Raman scattering \cite{gretarsson_two-magnon_2016}. Such strong coupling of HEs to magnons supports the idea that AFM correlations play a significant role in HD binding. Note that even though 3D long-range AFM order is lost above $T_N$, individual 2D layers continue to exhibit pronounced AFM correlations with a well-defined magnon spectrum in the paramagnetic phase since the scale of $J$ far exceeds $T_N$ \cite{gretarsson_persistent_2016, fujiyama_two-dimensional_2012, gretarsson_two-magnon_2016}. Since it is in-plane correlations that are critical to the stability of HEs \cite{clarke_particle-hole_1993, shinjo_density-matrix_2021, lenarcic_charge_2014, terashige_doublon-holon_2019}, this behavior is consistent with the absence of anomalies in $\tau$ near $T_N$.

To identify the specific intra-excitonic transition responsible for the 1.5 THz peak, we compare our data with many-body model calculations. Starting from the Hubbard model, we consider an extended single-band $t$-$J$ model on the 2D square lattice (Methods) that is described by the Hamiltonian: 

\begin{align}
    \begin{split}
        H_{tJV} &=t_{\textrm{NN}}\sum_{\langle ij\rangle,s}(h_{is}^\dagger h_{js} - d_{is}^\dagger d_{js}  + {\rm h.c.}) + U\sum_{i}n_{di} \\
        &- V\sum_{\langle ij\rangle}(n_{di}n_{hj}+n_{hi}n_{dj}) + J \sum_{\langle ij \rangle}\Big(\mathbf{S}_{i}\cdot\mathbf{S}_{j}-\frac{1}{4}\delta_{1,n_{i}n_{j}}\Big) \label{EqHtj_main}\\
    \end{split}
\end{align}

\noindent 
where $h_{is}^\dagger$ ($d_{is}^\dagger$) is the creation operator of a holon (doublon) defined using spin $s$ on site $i$, $n_{hi}=(1/2)\sum_{s}h_{is}^\dagger h_{is}$ and $n_{di}=(1/2)\sum_{s}d_{is}^\dagger d_{is}$ are the holon and doublon number operators respectively, $\mathbf{S}_{i}$ is the spin operator on site $i$, and the summation $\langle ij\rangle$ runs over nearest-neighbor pairs (Methods). The values of the nearest-neighbor hopping $t_{\textrm{NN}}$ = 0.26 eV, nearest-neighbor inter-site Coulomb energy $V$ = 0.39 eV and exchange $J = 4t_{\textrm{NN}}^2/U$ = 0.06 eV were chosen consistently with reported values for Sr$_{2}$IrO$_{4}$ \cite{watanabe_microscopic_2010, wang_twisted_2011, kim_magnetic_2012}, with $U$ being the on-site Coulomb energy. Eigenstates within the subspace of a single HD pair are calculated via exact diagonalization of $H_{tJV}$ on a 26-site square cluster using the Lanczos algorithm \cite{lenarcic_charge_2014, lenarcic_ultrafast_2013}. As shown in Figure 4a, the eigenvalue spectrum at zero center-of-mass momentum features four discrete bound HE levels spaced by several THz, separated from a higher energy HD continuum. The four HE wavefunctions have $s$-, $d$-, $s$-, and $p$-wave symmetries in order of lowest to highest energy, clearly departing from the hydrogenic series as previously predicted \cite{wrobel_excitons_2002, lenarcic_charge_2014, tohyama_symmetry_2006, huang_spin-mediated_2020}. These symmetries can be directly visualized in real space through the density correlator $D_j = \langle \psi^{hd}_m|n_{hj}n_{d0}|\psi^{hd}_m\rangle$, where $|\psi^{hd}_{m}\rangle$ is the HD pair wavefunction, which describes the probability of measuring a holon at site $j$ given a doublon at the origin for the four different excitonic states $m = 1 \rightarrow 4$ (Figure 4b). Whereas $D_j$ for the $s$- and $d$-states is 4-fold rotational symmetric, that for the $p$-state is only 2-fold symmetric.

Our numerical simulations confirm that the observed 1.5 THz peak lies within the predicted frequency scale of low-lying intra-excitonic excitations in Sr$_{2}$IrO$_{4}$, which is an order of magnitude smaller than $J$. A leading candidate is the dipole-allowed $s$- to $p$-state transition appearing at 2.85 THz in our calculations (Figure 4a). Since HEs are generated by excitation across the Mott gap in our experiments, a finite initial population of excited HE states is expected. Moreover, the proximity of the $p$-state to the HD continuum (Figure 4a, Extended Data Figure 4) together with strong coupling to magnons implies a shortened $p$-state lifetime, which may explain the broad linewidth of the observed 1.5 THz mode. We note that while the experimental detection of a single mode is not sufficient to pin down the excitonic spectrum, our theoretical results show that its energy scale is consistent with an intra-HE transition. 

To verify that $J$, in addition to $V$, contributes to binding the HEs identified in our simulations, we evaluated the deviation of the spin correlator relative to the AFM ground state for each HE state (Methods). The positions of the holon and doublon were fixed at their most probable location as determined by $D_{j}$ (Figure 4b).  As shown in Figure 4c, the HE states clearly perturb the underlying AFM order, with higher energy configurations correlated with a larger range of disturbance to the AFM background. 

Our tr-TDTS and theoretical results together establish that HEs can exist as metastable neutral quasiparticle excitations in a 2D AFM Mott insulator. Moreover, they demonstrate a pathway to prepare a HE fluid through photo-excitation resonant with the Mott gap. The energetic and dynamical properties of the HEs can in principle be controlled \textit{in situ} by tuning $J$ using mechanical or optical perturbations \cite{mentink_ultrafast_2015}, serving new technological applications while also further confirming their exchange-bound nature. More generally, these results suggest that 2D magnetic Mott insulators, which host myriad ordered and quantum disordered phases, are a promising platform for discovering novel excitonic states. \\

\noindent\textbf{Acknowledgements}

\noindent The authors thank Leon Balents, Mengxing Ye, Rick Averitt, Patrick Lee, Matteo Mitrano, Michele Buzzi, Scott K. Cushing, Peter Prelovšek, Denis Golež, and Victor Galitski for useful discussions. We thank S. J. Moon for sharing his equilibrium optical conductivity results. This work is supported by ARO MURI Grant No. W911NF-16-1-0361. D.H. acknowledges support for instrumentation from the David and Lucile Packard Foundation and from the Institute for Quantum Information and Matter (IQIM), an NSF Physics Frontiers Center (PHY-1733907). S.D.W. acknowledges partial support via NSF award DMR-1729489. X.L. acknowledges support from the Caltech Postdoctoral Prize Fellowship and the IQIM. M.B. acknowledges support from the Deutsche Forschungsgemeinschaft (DFG, German Research Foundation) under Germany’s Excellence Strategy Cluster of Excellence Matter and Light for Quantum Computing (ML4Q) EXC 2004/1 390534769. Z.L. was funded by the Gordon and Betty Moore Foundation’s EPiQS initiative, Grant No. GBMF4545, and J1-2463 project and P1-0044 program of the Slovenian Research Agency. \\

\noindent\textbf{Author Contributions Statement}

\noindent O.M. and D.H. conceived the experiment. O.M constructed the time-resolved time-domain THz spectrometer with contributions from X.L. and N.J.L. O.M, X.L., H.N, and Y.H. collected and analyzed the data. Z.L. performed the exact diagonalization calculations and interpreted the results with O.M and M.B. Z.P. and S.W. synthesized and characterized the sample. O.M. and D.H. wrote the manuscript with input from all authors. \\

\noindent\textbf{Competing Interests Statement} 

\noindent The authors declare no competing interests.\\

\newpage
\noindent\textbf{Methods}

\noindent\textbf{A. Sample Growth}

Large millimeter sized single crystals of Sr$_{2}$IrO$_{4}$ were grown using a self-flux technique from off-stoichiometric quantities of IrO$_{2}$, SrCO$_{3}$, and SrCl$_{2}$. The ground mixtures of powders were melted at 1470 \textdegree C in partially capped platinum crucibles. The soaking phase of the synthesis lasted for $>$20 h and was followed by a slow cooling at 2  \textdegree C/h to reach 1400 \textdegree C. From this point, the crucible is brought to room temperature through rapid cooling at a rate of 100 \textdegree C/h. The (001) face of the crystals were polished to a mirror finish using diamond lapping paper with a grit size of 1 $\mu$m.\\

\noindent\textbf{B. Time-Domain THz Spectroscopy}

    \noindent\textit{1. Experimental apparatus}
    
    The tr-TDTS setup was seeded by 800 nm, 35 fs pulses produced by a Ti:sapphire amplifier operating at 1 kHz, which was split into three arms. The first arm (3.5 mJ pulse energy) was sent into an optical parametric amplifier (OPA) that served as the tunable near-infrared (NIR) pump source. The second arm (0.6 mJ pulse energy) was used to generate the broadband THz-frequency probe through optical rectification of the 800 nm pulses. The third arm (1 $\mu$J pulse energy) was reserved for the electro-optic sampling (EOS) gate pulse used to measure the THz electric field $E(t_{\textrm{EOS}})$. The temporal delay $t_{\textrm{EOS}}$ between the 800 nm EOS gate pules and the THz probe was controlled with a motorized delay stage. 
    
    Two different configurations of detection and generation crystals were used in these measurements. The first utilized a 0.2 mm thick $<$110$>$ GaP crystal for generating the THz pulse and a 0.2 mm thick $<$110$>$ GaP crystal mounted on 1 mm thick $<$100$>$ GaP for EOS detection, yielding a bandwidth of 0.5 THz to 6 THz. The second utilized 1.0 mm thick $<$110$>$ ZnTe crystals for both generation and EOS detection, yielding a bandwidth of 0.35 THz to 2 THz. The entire tr-TDTS apparatus was enclosed in a N$_{2}$ gas purged environment.\\

    \noindent\textit{2. Determination of equilibrium optical constants}
    
    The TDTS setup was utilized in transmission geometry to measure the equilibrium optical constants of Sr$_{2}$IrO$_{4}$ shown in Figure 1c. The equilibrium complex index of refraction $\tilde{n}\left(\omega\right)$ was obtained by measuring the transmission of a THz pulse through the sample along the (001) axis. In this case, ZnTe crystals were used for both generation of the THz pulse and EOS detection. The THz electric field was first measured by transmitting through the sample sitting over an aperture. Then, the sample was removed from the aperture and the THz electric field was measured again, keeping all other parameters fixed. These two field transients were divided in the frequency domain to obtain the experimental complex transmission $\tilde{t}_{Exp.}\left(\omega\right)$. To extract $\tilde{n}\left(\omega\right)$, the difference between  $\tilde{t}_{Exp.}\left(\omega\right)$ and the expected theoretical response $\tilde{t}_{Th.}\left(\omega\right)=\ \frac{4\tilde{n}\left(\omega\right)}{\left[\tilde{n}\left(\omega\right)+1\right]^2}\times\left[\mathrm{exp}\frac{i\omega d}{c}\left(\tilde{n}\left(\omega\right)-1\right)\right]$  was minimized using a least-squares algorithm at each $\omega$ with $\tilde{n}\left(\omega\right)$ as the variable of interest. Here $d$ is the sample thickness and $c$ is the speed of light in vacuum. \\
    
    \noindent\textit{3. Time-resolved time-domain THz spectroscopy measurements}
    
        The TDTS setup was utilized in reflection geometry to measure the photo-induced changes to the optical spectrum of Sr$_{2}$IrO$_{4}$. The THz probe beam was $s$-polarized and incident onto the (001) sample surface at a 30$^{\circ}$ angle of incidence. The pump beam was normally incident onto the sample and was cross-polarized with the THz probe.  The transient pump-induced change to the THz electric field is defined as $\Delta E\left(t_{\mathrm{EOS}}, t \right) = E_{Pumped}\left(t_{\mathrm{EOS}}, t \right)-E_{Static}\left(t_{\mathrm{EOS}}\right)$, where $t$ is the relative delay between the pump pulse and the EOS gate pulse. To measure the pump-induced waveform $\Delta E\left(t_{\mathrm{EOS}}, t \right)$, $t$ is kept fixed and both the EOS gate pulse and the pump pulse are swept along the EOS time axis $t_{\mathrm{EOS}}$ using two motorized delay stages that are synchronized with each other. The transient spectra shown in Figures 1 and 2 are obtained by repeating this measurement scheme at a series of values for $t$. 

        The differential measurement of the pump-induced waveform $\Delta E\left(t_{\mathrm{EOS}}, t \right)$ is enabled by mechanically chopping the NIR pump pulse at half the repetition rate of the probe pulse. The mechanical chopper served as the reference trigger for a lock-in amplifier, which filtered the EOS signal from a balanced photodiode. Such a detection technique yields a “pump-on” minus “pump-off” detection of the THz electric field. This technique can be operated with either one or two choppers and lock-in amplifiers. In the latter case, the static electric field can be detected simultaneously with the transient changes, ensuring that any spectral artifacts due to long-term drift are eliminated. The cost is a reduction in the signal-to-noise ratio by half. We utilized both methods in our measurements depending on the strength of the signal; both yielded identical results. When the single-chopper method was used, the static THz electric field and its transient changes were measured sequentially at each $t$ to ensure that there were no spectral artifacts. 
        
        In Extended Data Figure 2, $\Delta E\left(t_{\mathrm{fixed}}, t \right)$ was measured at a fixed point in the time-domain THz pulse. This was done by anchoring the EOS gate pulse to a particular point $t_{\mathrm{EOS}} = t_{\mathrm{\mathrm{fixed}}}$ while $t$ was swept using the motorized delay stage in the pump path. While these fixed electric field pump-probe traces do not produce a frequency-resolved response, they provide frequency-integrated information about the dynamics when $t_{\mathrm{fixed}}$ is anchored to the peak of $E(t_{\mathrm{EOS}})$ \cite{poellmann_resonant_2015}. Importantly, these traces are over 100 times faster to acquire than the full mapping of $\Delta E(t_{\mathrm{EOS}}, t)$. Since the measured response is frequency-integrated over the available bandwidth, the ZnTe crystals were utilized to ensure that only the excitonic response, which overlaps almost exactly with the ZnTe bandwidth of 0.35 to 2.0 THz, is captured with minimal influence from higher-energy features such as the phonon resonances.\\

    \noindent\textit{4. Determination of the transient optical conductivity }
    
    The transient changes to the index of refraction $\Delta \tilde{n}(\omega)$, and consequently the optical conductivity, can be determined from $\frac{\Delta \tilde{E}(\omega)}{\tilde{E}(\omega)}$ since it is equal to $\frac{\Delta \tilde{r}(\omega)}{\tilde{r}(\omega)}$, which depends on $\Delta\tilde{n}(\omega)$ \cite{hunt_manipulating_2015}. We use the thin film approximation \cite{hunt_manipulating_2015} in which the pump-induced portion of the sample can be considered as a thin film on top of the un-excited bulk with a homogeneous index of refraction $\tilde{n}^{\prime}\left(\omega\right)=\tilde{n}\left(\omega\right)+\Delta n(\omega)$. This approximation is justified (Supplementary Information Section III) since the THz pulse fully transmits through the 100 $\mu$m thick sample, while the pump pulse has a penetration depth of 73 nm \cite{bhandari_electronic_2019}, leading to a pump-probe penetration depth mismatch of greater than 3 orders of magnitude. 
    
    An electrodynamic analysis of this situation using the  Maxwell’s equations yields an analytic solution for the transient changes to the optical conductivity \cite{hunt_manipulating_2015}: 

    \begin{equation}
        \Delta\tilde{\sigma}(\omega)= \left(\frac{1}{377\times d}\right)\frac{\frac{\Delta \tilde{E}(\omega)}{\tilde{E}(\omega)}\left(\tilde{n}^{2}(\omega)-1\right)}{\frac{\Delta \tilde{E}(\omega)}{\tilde{E}(\omega)}\left[\cos(\theta_{0})-\sqrt{\tilde{n}^{2}(\omega)-\sin^{2}(\theta_{0})}\right]+2\cos(\theta_{0})}
    \end{equation}
    where $d$ is the penetration depth of the pump pulse and $\theta_{0}$ is the angle of incidence. \\

\noindent\textbf{C. Drude-Lorentz Fitting}

Both the real and imaginary parts of the transient optical conductivity, $\Delta\sigma_{1}(\omega)$ and $\Delta\sigma_{2}(\omega)$, were simultaneously fit at each $t$ with the following Drude-Lorentz model: 

\begin{equation}
\frac{D}{2}\left[\frac{1}{\gamma_{\textrm{Drude}}-i\omega}\right]+\frac{L_{\mathrm{HE}}}{2}\left[\frac{\omega}{i\left({\omega_{\mathrm{HE}}}^2\ -\omega^2\ \right)+\omega \gamma_{\mathrm{HE}}}\right]+\frac{L_{\mathrm{Bgd.}}}{2}\left[\frac{\omega}{i\left({\omega_{\mathrm{Bgd.}}}^2\ -\omega^2\ \right)+\omega \gamma_{\mathrm{Bgd.}}}\right]
\end{equation}\\
The first term is a Drude term. The second term is a Lorentzian term representing the HE mode. The final Lorentzian term describes a weak background, likely caused by pump-induced changes of higher energy features such as the phonon transitions. One or two background Lorentzians were used depending on the dataset. The fitting constants $D$, $L_{\textrm{HE}}$, and $L_{\textrm{Bgd.}}$ are the strengths of the Drude, HE, and background terms, respectively. The fitting constants $\gamma_{\textrm{Drude}}$, $\gamma_{\textrm{HE}}$, and $\gamma_{\textrm{Bgd.}}$ are the widths of the Drude, HE, and background terms, respectively. The fitting constants $\omega_{\textrm{HE}}$ and $\omega_{\textrm{Bgd.}}$ are the central frequencies of the HE and background terms, respectively. For $t \leq 450$ fs, the value of $\gamma_{\textrm{Drude}}$ was left as a free parameter, while for later $t$ it was fixed to its average fitted value in order to constrain the number of free parameters and improve the quality of the fit. A high quality of fit is achieved as shown in Figure 2a-f. The spectral weights in Figure 2g are found by integrating over each term in the fit individually. In Extended Data Figure 1, a representative dataset is shown along with each of the individual components of the fit.\\

\noindent\textbf{D. Exponential Fitting}
To fit the data in Figure 2g, Figure 3a, Extended Data Figure 2, and Extended Data Figure 3, we used an exponential function of the form: 

\begin{equation}
    \frac{1}{2}\left[1+\ \mathrm{erf}\left(2\sqrt2\ \left(\frac{t-t_0}{t_r}\right)\ \right)\right]\times\left[\sum_{i}\left[A_i\ \mathrm{exp} \left(-\frac{t-t_0}{\tau_i}\right)\right]+b\right]
\end{equation}\\
\noindent
where $t$ is the pump-probe time delay, and the fitting parameters $t_{0}$, $t_{r}$, $b$, and $A_{i}$ and $\tau_{i}$ are time zero, the rise time, the pump-induced offset, and the strength and decay constant of each exponential $i$, respectively. The SW data in Figure 2g, the temporal cuts of $\Delta\sigma_{1}(\omega)$ in Figure 3a, and the $\Delta E/E$ data in Extended Data Figure 2 were fit using $i$ = 1, while the $\Delta R/R$ traces in Extended Data Figure 3 were fit with $i$ = 2. \\

\noindent\textbf{E. Transient Reflectivity Measurements}
The measurement apparatus was seeded by a Ti:sapphire regenerative amplifier operating at 100 kHz generating 100 fs pulses centered at 800 nm. A portion of the output served as the source of the incident probe beam, while the remaining power seeded an optical parametric amplifier that was used to generate the 0.6 eV pump. The pump beam was linearly polarized and passed through a 5 kHz mechanical chopper before impinging onto the sample at normal incidence. The pump-probe delay was controlled by a motorized delay stage, and the pump induced changes to the reflectivity were measured with lock-in detection. 

A high quality of fit using a double exponential formula (Methods) is achieved as shown in Extended Data Figure 3. The decay constant of the first exponential $\tau_{1}$ gives a near temperature independent value of roughly 130 fs, and accounts for intraband scattering consistent with our own tr-TDTS measurements and previous time-resolved reflectivity studies \cite{hsieh_observation_2012, dean_ultrafast_2016}. The second timescale, $\tau_{2}$, accounts for interband decay \cite{okamoto_photoinduced_2011, hsieh_observation_2012} and is the value plotted in Figure 3b as a function of temperature.\\

\noindent\textbf{F. Numerical Calculations}
Sr$_2$IrO$_4$ is well-described theoretically by the single-band Hubbard model \cite{kim_novel_2008, wang_twisted_2011} in the presence of the nearest-neighbor (NN) Coulomb interaction: 
\begin{equation}
H= -  t_{NN} \sum_{ \langle ij \rangle s} (c^\dagger_{j s}  c_{i s} + {\rm h.c.})   
 + U \sum_i n_{i \uparrow} n_{i \downarrow}, 
+ V \sum_{\langle ij\rangle} n_{i} n_{j}.
 \label{hub}
\end{equation}

\noindent
The sum runs over NN pairs of sites $\langle ij\rangle$ on the 2D square lattice and $c_{js}$ are the fermionic annihilation operators for the electron with spin $s=\pm \frac{1}{2}$ on site $j$. Before the arrival of the pump pulse, the half-filled system is in its ground state. The pump pulse excites an electron across the Mott gap, creating a holon in the LHB and a doublon in the UHB. It is known that through intraband cooling, the photo-excited holons and doublons quickly ($\sim$ 60 fs) relax to the band minimum \cite{hsieh_observation_2012, dean_ultrafast_2016, piovera_time-resolved_2016, okamoto_photoinduced_2011}. However, further relaxation via recombination across the Mott gap is bottlenecked by the large amount of energy ($\sim$ 0.5 eV) that needs to be transferred to other bosonic degrees of freedom (spins, phonons) with much smaller energy scales ($\sim$ 1-100  meV), leading to typical timescales exceeding 1 ps \cite{okamoto_photoinduced_2011, hsieh_observation_2012}. Because intraband relaxation and recombination occur on drastically different timescales, we canonically transform the Hubbard model so as to separate them evidently in different orders of the small $1/U$ parameter \cite{lenarcic_charge_2014, lenarcic_ultrafast_2013}. While the exact canonical transformation preserves the model, we drop the terms that appear with parametrically small prefactors ($1/U^{n}, n>2$) or are expected to be subleading compared to terms already included in lower orders of the $1/U$ expansion. We end up with an effective model $\tilde H=H_{tJV} + H_{rc}$, where $H_{tJV}$ captures holon and doublon hopping, on-site and NN Coulomb interaction, and the interaction-mediated spin exchange, while $H_{rc}$ captures the recombination processes that can be treated as a perturbation:

\begin{equation}
     \tilde H=H_{tJV}+H_{rc}\label{EqHtilde}
\end{equation}

\begin{align}
    \begin{split}
        H_{tJV} &=t_{\textrm{NN}}\sum_{\langle ij\rangle,s}(h_{is}^\dagger h_{js} - d_{is}^\dagger d_{js}  + {\rm h.c.}) + U\sum_{i}n_{di} \\
        &- V\sum_{\langle ij\rangle}(n_{di}n_{hj}+n_{hi}n_{dj}) + J \sum_{\langle ij \rangle}\Big(\mathbf{S}_{i}\cdot\mathbf{S}_{j}-\frac{1}{4}\delta_{1,n_{i}n_{j}}\Big) \label{EqHtj}\\
    \end{split}
\end{align}

\begin{equation}
    H_{rc}=t_{rc}\sum_{(ijk),ss'}\left(h_{ks}d_{js'}\vec{\sigma}_{s\bar{s}'}\cdot\mathbf{S}_{i} + {\rm h.c.} \right),
    \quad t_{rc}=\frac{2t^2}{U}=\frac{J}{2}. \label{EqHrec}
\end{equation}\\
\noindent
We have introduced holon and doublon creation operators,
$h_{is}^\dagger=c_{is}(1-n_{i\bar s}), \quad  
d_{is}^\dagger=c_{i\bar s}^\dagger n_{is}$, 
and corresponding density operators 
$n_{hi}=(1/2)\sum_{s}h_{is}^\dagger h_{is}$,
$n_{di}=(1/2)\sum_{s}d_{is}^\dagger d_{is}$. Here $\bar{s}=-s$ is the opposite spin of $s$ and $\vec\sigma=\{\sigma^x,\sigma^y,\sigma^z\}$ is a vector of Pauli matrices so that in the above notation $\sigma_{s,s'}^{a}$ corresponds to the $(\frac{3}{2}-s, \frac{3}{2}-s')$ component of the $\sigma^{a}$ matrix. The sum over $(ijk)$ runs over $i,j,k$, where $j\neq k$ are the NN sites to site $i$. 

Motivated by the fluence independent tr-TDTS response (Extended Data Figure 2), indicating the irrelevance of exciton-exciton interactions, we assume low densities of photo-excited HD pairs. Having already separated sectors with different number of HD pairs on the level of the Hamiltonian - $H_{tJV}$ conserves the number of HD pairs while $H_{rc}$ perturbatively changes it - we can extract the metastable states at the bottom of the UHB using the Lanczos algorithm for exact diagonalization of $H_{tJV}$ within the sector with one HD pair, while neglecting $H_{rc}$. These states are ostensibly the terminal point of the intraband relaxation process and, as we will point out with several indicators, correspond to excitonic states of bound HD pairs that lead to the formation of the transient insulating phase. The Lanzcos approach is ideal for this problem because it is suited for the calculation of the lowest eigenstates within the sector with one HD pair. Moreover, the Lanczos approach allows us to treat the system using periodic boundary conditions on $N=26$ sites, which is larger than what would be possible if we had considered the full exact diagonalization of the Hubbard model directly. We use $N_{Lan}=160,180$ Lanczos basis vectors for which the lowest eigenstates are well converged. For the $N=20$ case shown in Figure S10, $N_{Lan}=120,140,160$ is used.

In Figure 4 of the main text, we showed three main pieces of evidence for the formation of excitonic states. First, in Figure 4a, it is shown that several states are well-separated with finite gaps from a more densely-spaced set of states that represents the free HD continuum. Second, in Figure 4b, for each of these states, we use the density correlator $D_j = \langle \psi^{hd}_{m}|n_{hj}n_{d0}|\psi^{hd}_{m}\rangle$, which gives the probability of measuring a holon at site $j$ given a doublon fixed at the origin. $|\psi_m^{hd}\rangle \equiv |\psi^{hd}_{\vec{k}=[0,0],m}\rangle$ is the HD pair wavefunction of the $m$-th eigenstate in the single HD pair sector with zero center-of-mass momentum. It is apparent that the holon is well-localized around the doublon, indicating the formation of a bound state. Finally, in Figure 4c, we display the deviations in the spin correlator of the excitonic states relative to the AFM ground state in the spin sector without HD pairs, defined as $\delta\langle\textbf{S}_{i}\cdot\textbf{S}_{j}\rangle = \frac{\langle\textbf{S}_{i}\cdot\textbf{S}_{j}\rangle_{\textrm{HD}} - \langle\textbf{S}_{i}\cdot\textbf{S}_{j}\rangle_{\textrm{AFM}}}{\langle\textbf{S}_{i}\cdot\textbf{S}_{i+1}\rangle_{\textrm{AFM}}}$. This quantity gives the relative deviations in the spin bond energy as it is normalized by the spin bond energy in the AFM ground state. For this calculation, we fix the holon and doublon to their most probable locations as determined by the value of $D_{j}$. It is clear that the presence of the HD pair disrupts the AFM motif, and that the extent of this disruption becomes larger for the higher-energy excitonic states. This behavior supports the notion of a spin-mediated binding mechanism.  

In order to determine the symmetry of these excitonic states, we calculate the matrix elements $|\langle \psi^{hd}_{\vec{k},m} | O_{sym} | \psi^{hd}_{\vec{k},1} \rangle |^2$ at a fixed center-of-mass momentum $\vec{k}$ from the lowest eigenstate $\ket{\psi_{\vec{k},1}^{hd}}$ to the excited states $\ket{\psi^{hd}_{\vec{k},m}}$ with respect to operators with a different (sym=$\{s, p, d\}$) symmetry, for example,
\begin{align}
O_{s} &= \sum_{i} d_{i\rightarrow}^\dagger d_{i} 
	+ d_{i\leftarrow}^\dagger d_{i}
	+ d_{i\uparrow}^\dagger d_{i}
	+ d_{i\downarrow}^\dagger d_{i}, \\
O_{d} &= \sum_{i} d_{i\rightarrow}^\dagger d_{i} 
	+ d_{i\leftarrow}^\dagger d_{i}
	- d_{i\uparrow}^\dagger d_{i}
	- d_{i\downarrow}^\dagger d_{i}, \\	
O_{p} &= \sum_{i} d_{i\rightarrow}^\dagger d_{i} 
	- d_{i\leftarrow}^\dagger d_{i}.			
\end{align}
where $d_{i}=\sum_{s} d_{is}$ and $i\rightarrow / i\leftarrow / i\uparrow / i\downarrow$ are right/left/top/bottom neighboring sites of site $i$.
Depending on the symmetry of both states, the matrix element will be zero or finite. 
The lowest excitonic state at $\vec{k}=[0,0]$ has $s$-wave symmetry \cite{tohyama_symmetry_2006, lenarcic_ultrafast_2013, terashige_doublon-holon_2019}. From the above procedure, we can determine that the other excitonic states at $\vec{k}=[0,0]$ also have a definite symmetry: $d$-, $s$-, and $p$-wave, listed from lowest to highest energy respectively (Figure 4a, Extended Data Figure 4). States at the other high symmetry points also demonstrate a definite symmetry (Supplementary Information Section V). We should note that for Figure 4a and Extended Data Figure 4 in the main text, the dense continuum of states is obtained by superposing eigenstates calculated from diagonalization of the $H_{tJV}$ in the Krylov subspace, spanned starting from $O_{sym} | \psi^{hd}_{\vec{k},1} \rangle$ for different $sym=\{s,p,d\}$ and two choices of numbers of Lanczos basis vectors $N_{Lan}=160,180$. A standard Lanczos diagonalization of $H_{tJV}$ would yield a much sparser energy spectrum. Even so, we cannot confidently conclude if the $p$-wave exciton at $\vec{k}=[0,0]$ lies below the continuum at all momenta, despite its presence at $\vec{k}=[0,0]$. Thus, it is potentially unstable against decay into the continuum; see also \cite{shinjo_density-matrix_2021}. 

We should also note that the relative energies of the excitons of different symmetries at $\vec{k}=[0,0]$ depends on both $J$ and $V$ parameters, signaling the importance of both of these interactions. Furthermore, on small lattices, the order of states can also be affected by finite size effects. However, despite these dependencies on the strength of the microscopic parameters and the size of the lattice, our main conclusions are robust (Supplementary Information Section V). Regardless of the size and the specific choices of $J$ and $V$, there are always a series of even parity excitonic states below the odd parity $p$-wave state. Thus, we always find multiple optically-allowed transitions between various excitonic energy levels that occur at fractions of the overall binding energy. \\

\noindent\textbf{Data Availability}

Source data are provided with this paper. All other data that support the plots within this paper and other findings of this study are available from the corresponding author upon reasonable request.

\newpage
\noindent\textbf{Main Text Figures}\\
\\
\begin{figure}[h!]
\includegraphics[width=\textwidth]{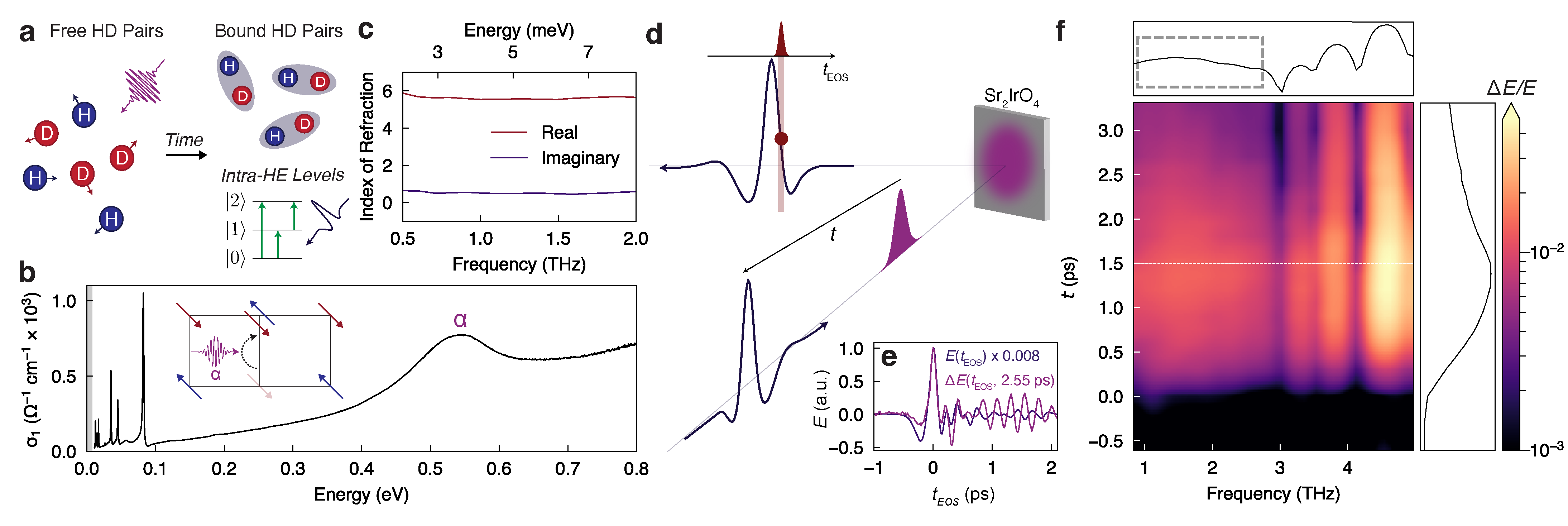}
\label{Fig1}
\end{figure}\\
\noindent\textbf{Figure 1 $|$ Electrodynamic properties of Sr$_2$IrO$_{4}$.} \textbf{a,} A photo-doping induced HD plasma transforms into a fluid of bound HEs as a function of time. Transitions between internal HE levels (green arrows), which are populated upon HD plasma decay, are probed by a THz pulse. \textbf{b,} Optical conductivity of Sr$_{2}$IrO$_{4}$  \cite{seo_infrared_2017}. Inset: Real-space depiction of HD creation via photo-excitation that is resonant with the Mott gap $\alpha$. \textbf{c,} Real and imaginary parts of the index of refraction of Sr$_{2}$IrO$_{4}$ in the THz regime (shaded gray energy window in panel b). \textbf{d,} Schematic of the tr-TDTS setup. The sample is excited by an intense near-infrared pulse (magenta) resonant with the $\alpha$ transition. The transient response at a fixed time-delay $t$ is then probed by a weak broadband THz pulse (blue). An 800 nm pulse (maroon) measures pump-induced changes of the reflected THz pulse through EOS (Methods). \textbf{e,} Equilibrium (blue) THz pulse $E(t_{\textrm{EOS}})$ and its pump induced change (magenta) $E(t_{\textrm{EOS}}, t=2.55$ ps). \textbf{f,} Differential change in the reflected THz spectrum of Sr$_{2}$IrO$_{4}$ taken at 80 K in response to $\alpha$ peak-resonant photo-excitation with fluence 2 mJ/cm$^{2}$. Top: Spectrum at $t$ = 1.5 ps. Gray box highlights the broad peak. Right: Frequency-integrated response as a function of $t$.\\

\newpage
\begin{figure}[h!]
\includegraphics[width=\textwidth]{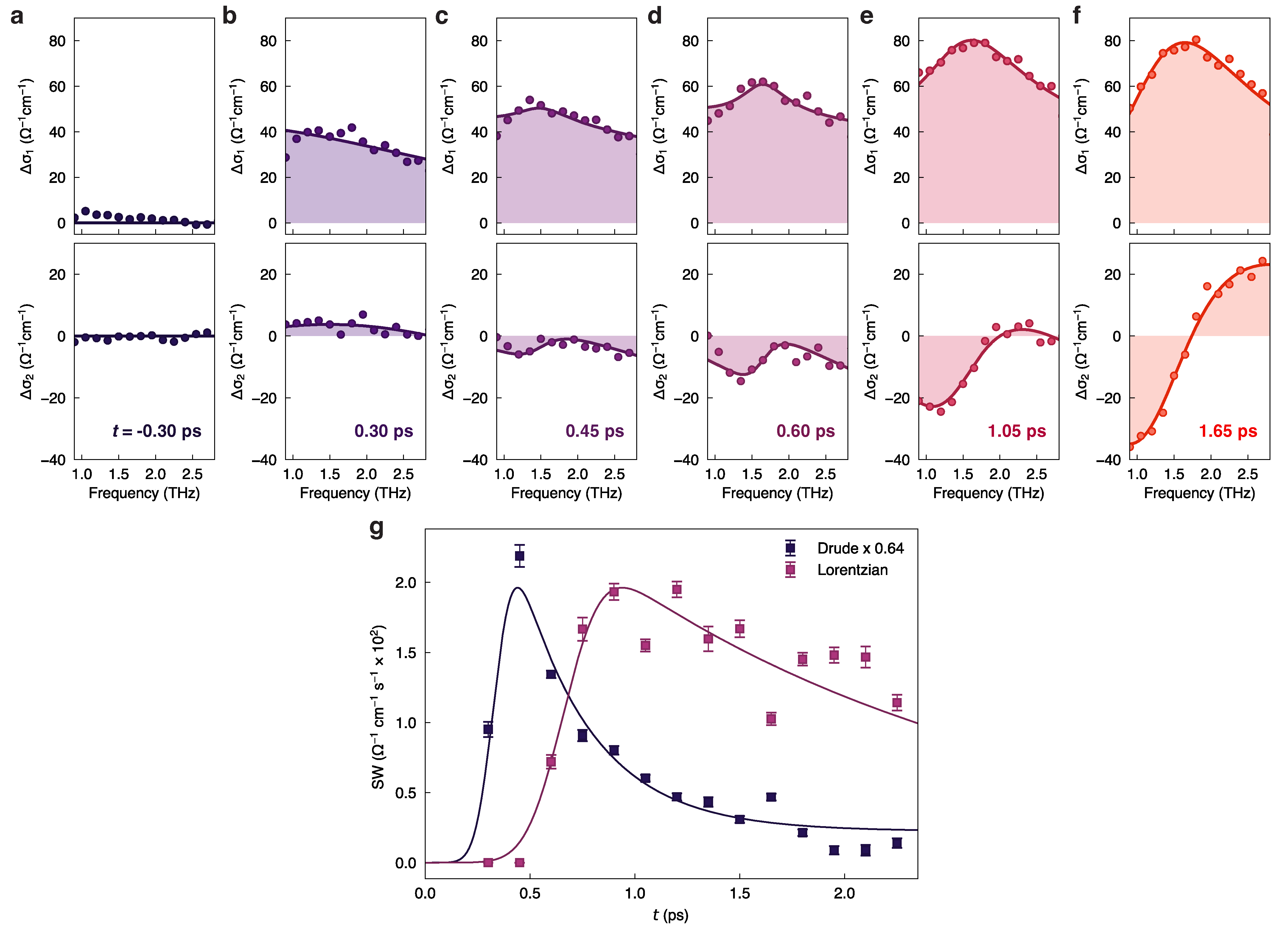}
\label{Fig2}
\end{figure}
\noindent
\textbf{Figure 2 $|$ Photo-doping induced optical conductivity transients of Sr$_{2}$IrO$_{4}$.} \textbf{a-f} $\Delta\sigma_{1}(\omega)$ (top panels) and $\Delta\sigma_{2}(\omega)$ (bottom panels) extracted from differential THz spectra at various $t$. Fits to the Drude-Lorentz model (Methods) are displayed as solid lines. \textbf{g,} SW of the Drude and HE Lorentzian terms versus $t$. The solid lines are fits to an exponential function (Methods). Error bars are obtained from the standard deviation of the least-squares-fitting algorithm.\\

\newpage
\begin{figure}
\includegraphics[width=90mm]{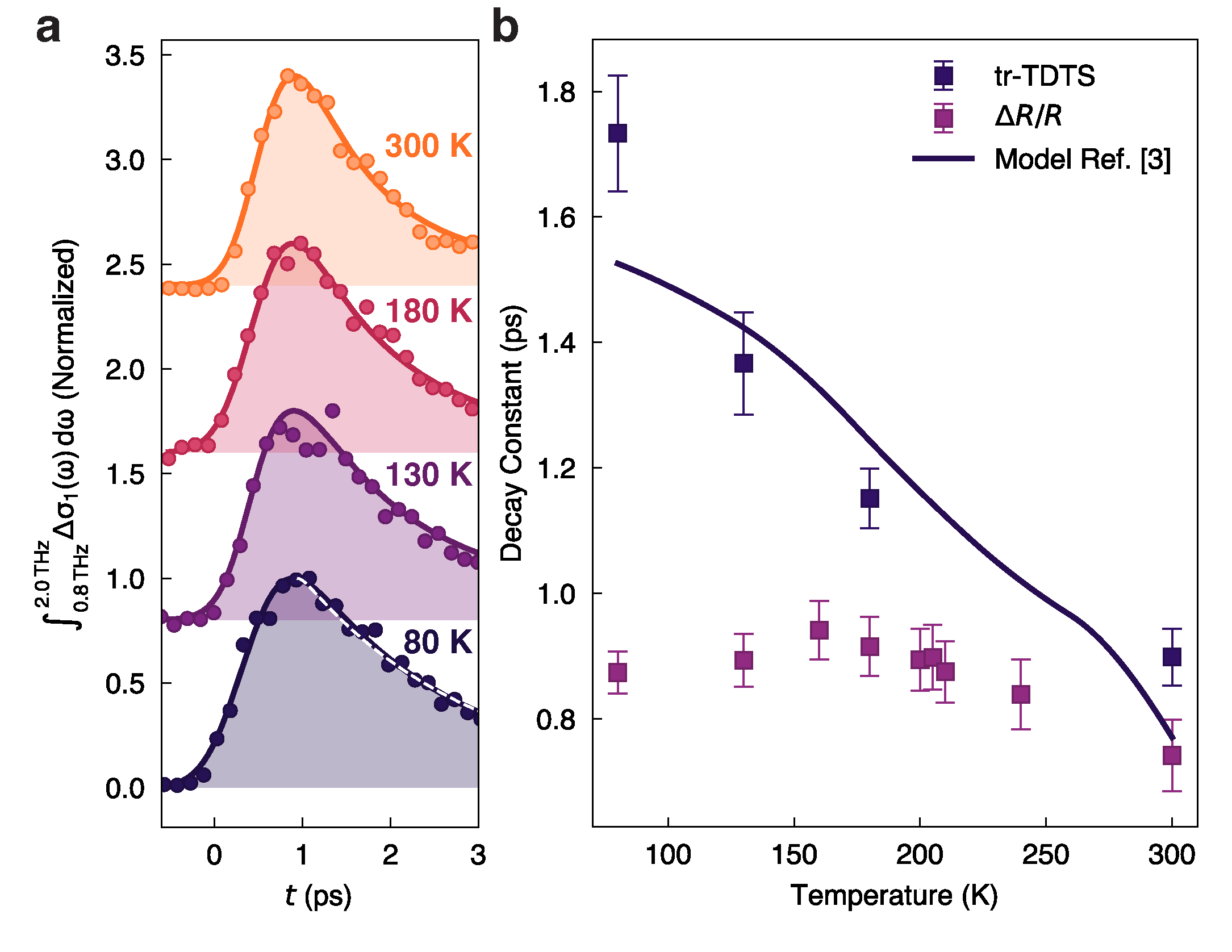}
\label{Fig3}
\end{figure}
\noindent\textbf{Figure 3 $|$ Temperature dependence of HE decay.} \textbf{a,} 
Temperature dependence of the photo-induced changes to $\Delta\sigma_{1}(\omega)$ integrated from 0.8 THz to 2.0 THz. Solid lines are fits to an exponential function (Methods). The white dashed line in the 80 K dataset is the exponential fit to the Lorentzian SW shown in Figure 2g, showing excellent agreement. \textbf{b,} Temperature dependence of the exponential decay constants extracted from the data in panel a and from the infrared reflectivity transients ($\Delta R/R$) shown in Extended Data Figure 3 (Methods). The solid line is a fit to the multi-magnon emission model described in Ref. [3]. Error bars are the standard deviation from the least-squares-fitting algorithm.

\newpage
\begin{figure}
\includegraphics[width=135mm]{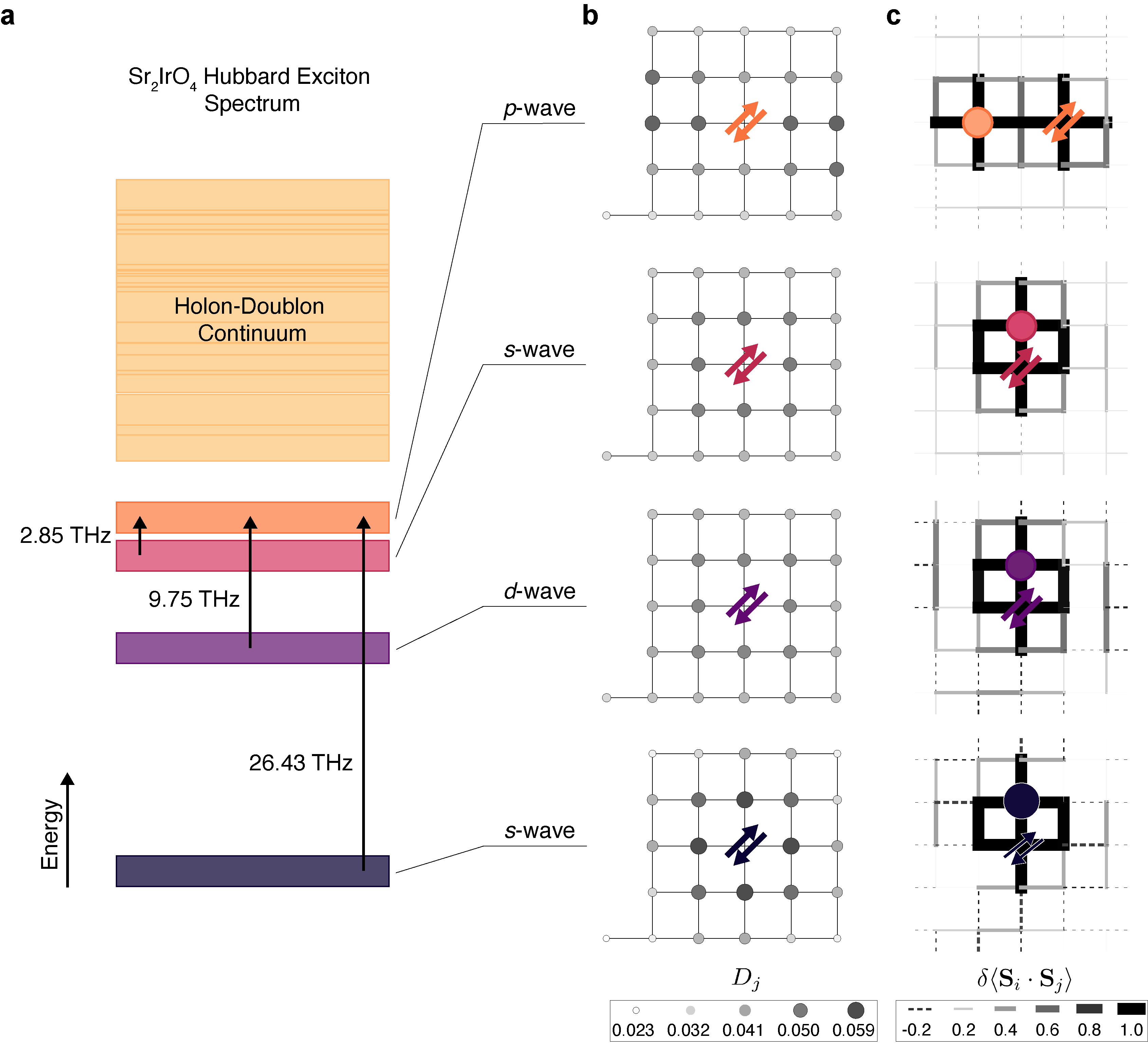}
\label{Fig4}
\end{figure}
\noindent\textbf{Figure 4 $|$ HE spectrum and characteristics obtained from effective model numerics.} \textbf{a,} Eigenstates of $H_{tJV}$ in the sector of a single HD pair at zero center-of-mass momentum calculated via exact diagonalization (see main text). Vertical arrows mark the optically-allowed internal HE transitions. \textbf{b,} The real space HD distribution on a square lattice represented through the density correlator $D_{j}=\langle\psi_{m}^{hd}\lvert n_{hj }n_{d0}\rvert \psi_{m}^{hd} \rangle$ for each of the four excitonic states labeled by $m$. The doublon is fixed at the center. The size and shade of each point indicate the strength of $D_{j}$, and therefore the probability of finding the holon at site $j$. \textbf{c,} Relative deviation of the spin correlator from the Heisenberg ground-state with the holon and doublon fixed at their most probable locations (Methods). The size and shade of each bond indicate the deviation normalized by the AFM ground state value, with dashed lines representing negative changes.

\newpage
\noindent\textbf{Extended Data Figures}\\
\\
\begin{figure}[h!]
\includegraphics[width=90mm]{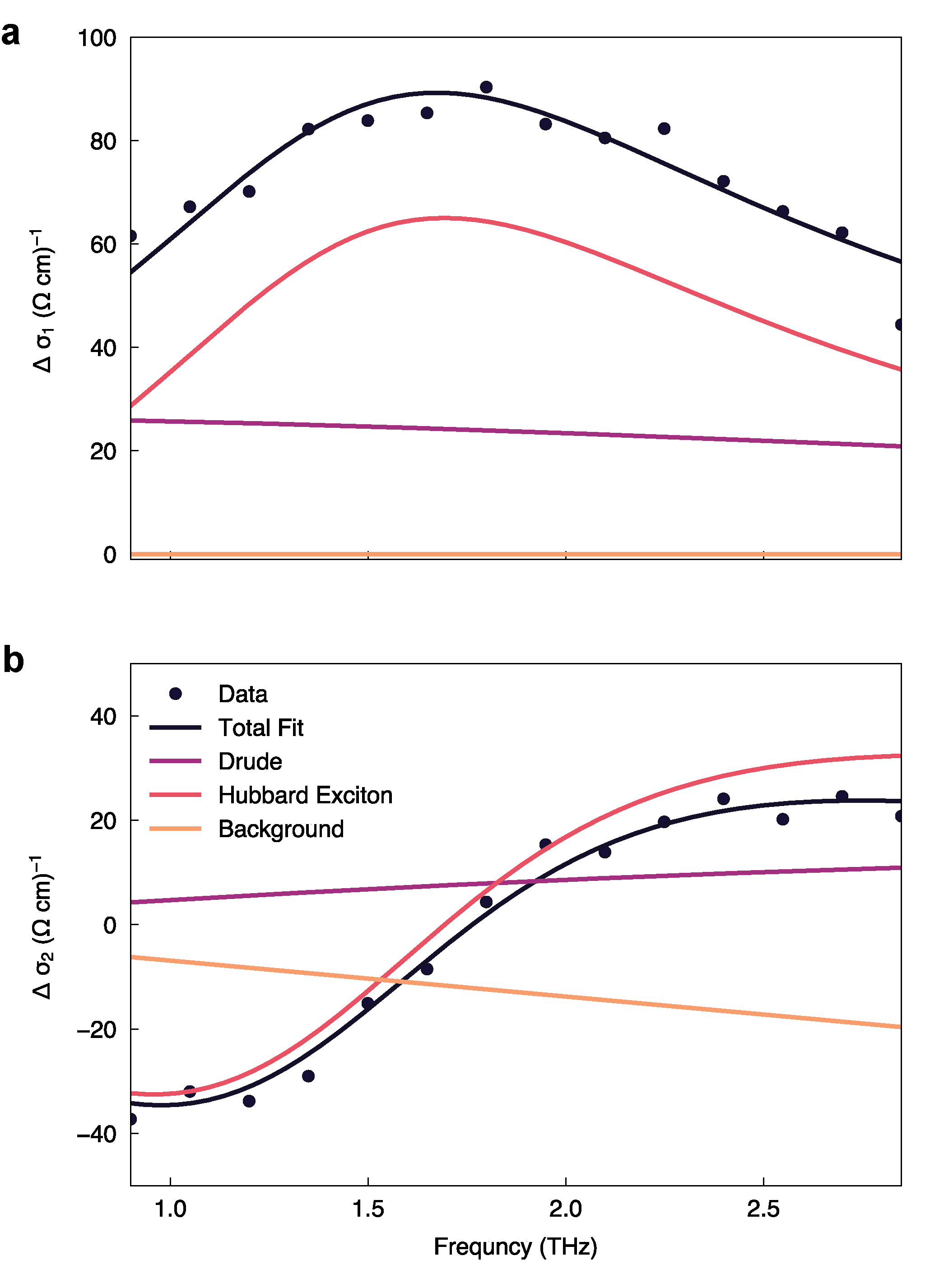}
\label{ExFig1}
\end{figure}\\
\noindent
\textbf{Extended Data Figure 1 $|$ Drude-Lorentz fitting of a typical tr-TDTS spectrum of Sr$_{2}$IrO$_{4}$.} Individual components of the Drude-Lorentz fitting of \textbf{a,} $\Delta\sigma_{1}(\omega)$ and \textbf{b,}  $\Delta\sigma_{2}(\omega)$ plotted with the original data at $t = 1.5$ ps.  The pump energy was fixed at 0.6 eV, resonant with the $\alpha$ transition, and set to a fluence of 2 mJ/cm$^{2}$. The temperature of the sample was 80 K.

\newpage

\begin{figure}
\includegraphics[width=90mm]{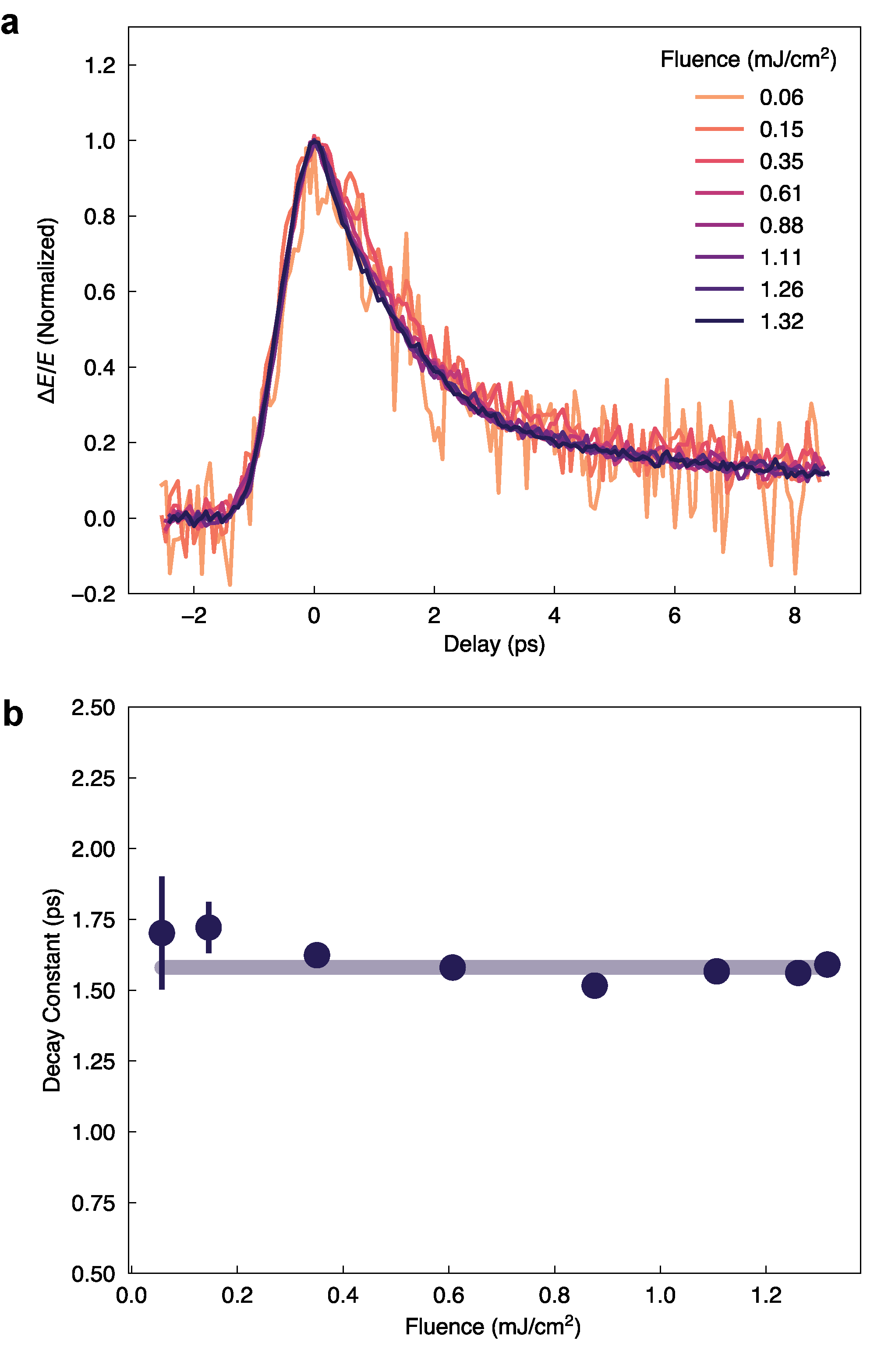}
\label{ExFig2}
\end{figure}
\noindent
\textbf{Extended Data Figure 2 $|$ Fluence independence of THz frequency decay dynamics.} \textbf{a,} $\Delta E$($t_{\textrm{EOS}}$, $t$) traces taken with $t_{\textrm{EOS}}$ fixed to the time where $E$($t_{\textrm{EOS}}$) is maximal (Methods) plotted as a function of the fluence of the $\alpha$-resonant (0.6 eV) photo-excitation. Data was collected at a sample temperature of 80 K. \textbf{b,} Decay constants extracted from an exponential fitting (Methods) of the traces in panel a. The solid line is a guide to the eye. Error bars are the standard deviation from the least-squares-fitting algorithm. 
\newpage

\begin{figure}
\includegraphics[width=90mm]{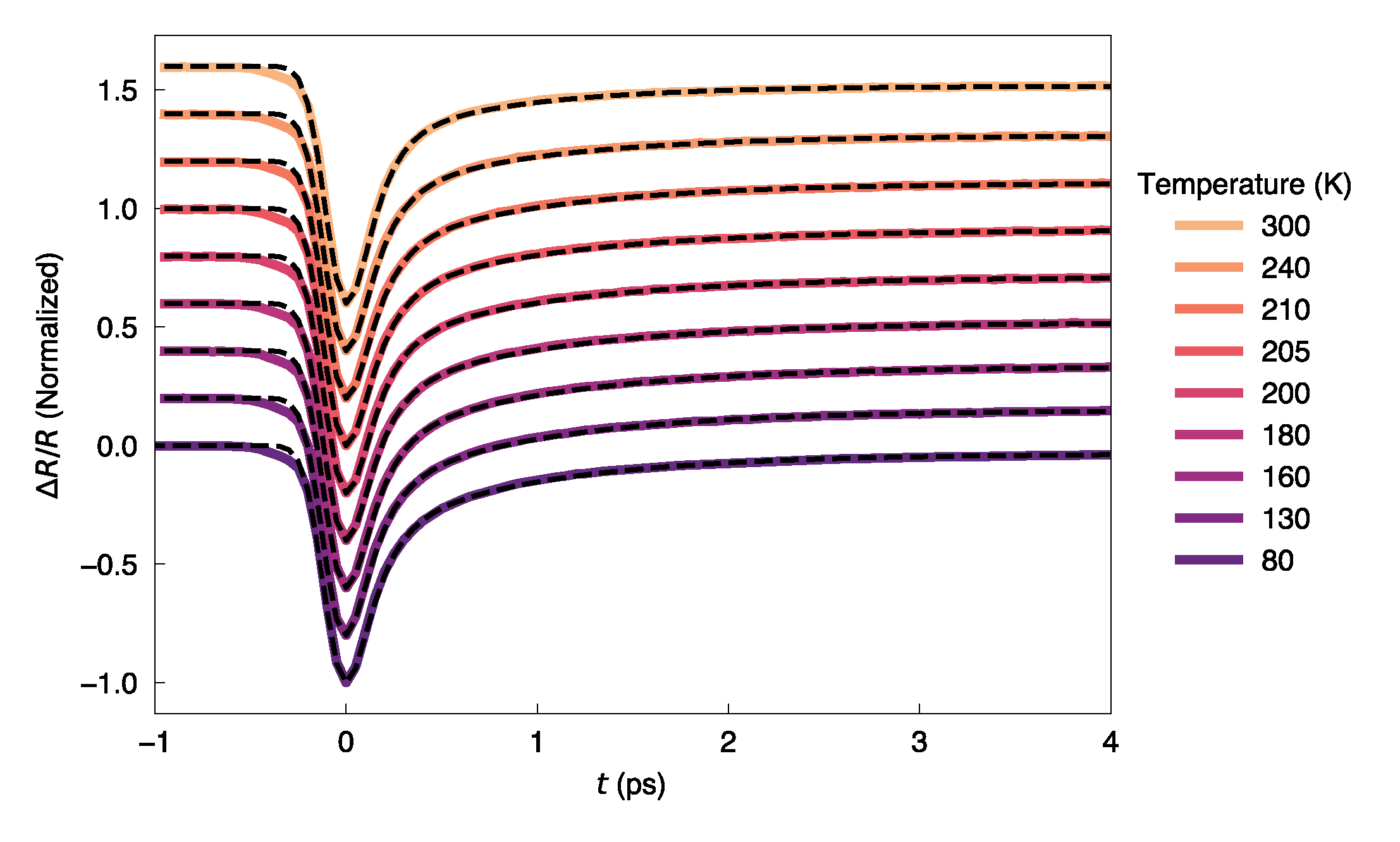}
\label{ExFig3}
\end{figure}
\noindent
\textbf{Extended Data Figure 3 $|$ Temperature dependence of differential infrared reflectivity.} $\Delta R/R$ traces taken on Sr$_{2}$IrO$_{4}$ as a function of temperature. The probe energy was fixed at 1.55 eV. The pump energy was fixed at 0.6 eV, resonant with $\alpha$, and a fluence near 2 mJ/cm$^{2}$ was used. The black dashed lines are fits to a double exponential function (Methods).

\newpage

\begin{figure}
\includegraphics[width=90mm]{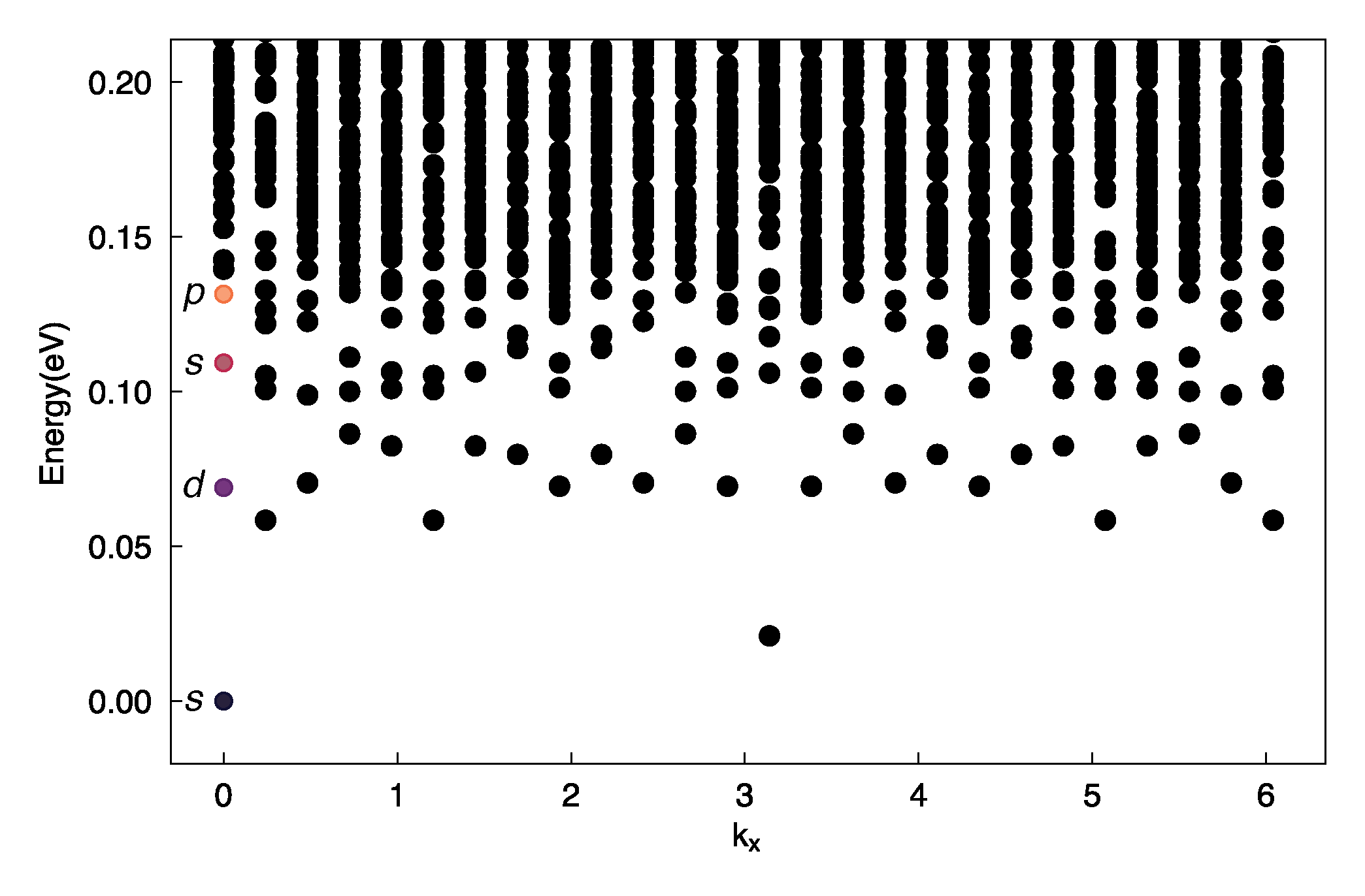}
\label{ExFig4}
\end{figure}
\noindent
\textbf{Extended Data Figure 4 $|$ Simulated dispersion of the HE states.} Eigenenergies of $H_{tJV}$ in the sector of a single HD pair projected onto the $k_{x}$ axis throughout the entire reciprocal space of the 26 site lattice. At zero center-of-mass momentum, $\vec{k}=[0,0]$, the four lowest states are colored to indicate which excitonic states are depicted in Figure 4 of the main text. We define 0 eV to be the energy of the lowest-energy state.
\newpage

\newpage
%
%


\widetext

\begin{center}
\textbf{\large Supplementary Information}
\end{center}
\setcounter{equation}{0}
\setcounter{figure}{0}
\setcounter{table}{0}
\setcounter{page}{1}
\makeatletter
\renewcommand{\theequation}{S\arabic{equation}}
\renewcommand{\thefigure}{S\arabic{figure}}
\renewcommand{\bibnumfmt}[1]{[S#1]}
\renewcommand{\citenumfont}[1]{S#1} 

\noindent\textbf{I. ZnTe-based tr-TDTS data on Sr$_{2}$IrO$_{4}$}

We performed additional time-resolved time-domain THz spectroscopy experiments using a ZnTe-based spectrometer (Methods). This spectrometer has a bandwidth of 0.35 THz to 2 THz, allowing us to probe at lower energies compared to GaP. As shown in Figure S1, we found no additional features below the finite energy peak centered at $\sim$ 1.5 THz.

\begin{figure}[h]
    \includegraphics[width=50mm]{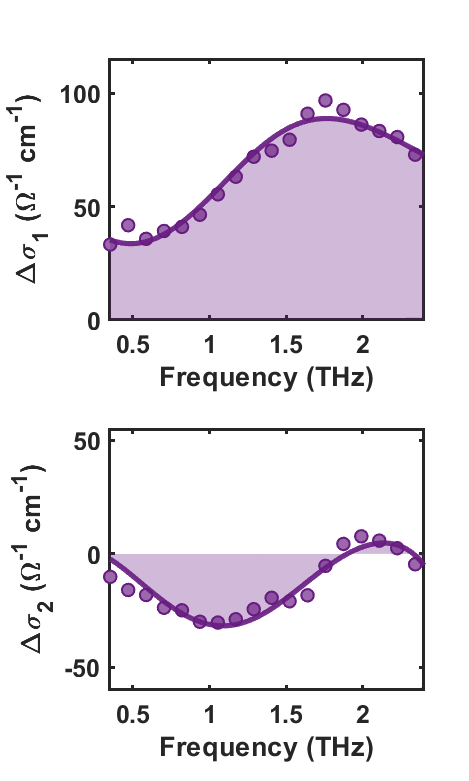}
    \label{ZnTeData}
    \caption{\textbf{Photo-doping induced optical conductivity transients of Sr$_{2}$IrO$_{4}$ probed with lower-frequency THz pulses.} Real (top panel) and imaginary (bottom panel) parts of the pump-induced change of optical conductivity extracted from differential THz spectra measured using a ZnTe-based spectrometer. The pump was tuned to 0.6 eV (resonant with the $\alpha$ transition) and a fluence of 2.0 mJ/cm$^{2}$. The data was taken at 80 K. Fits to the Drude-Lorentz model (Methods) are displayed as solid lines.}
    
\end{figure}

\newpage
\noindent\textbf{II. Ruling out alternative interpretations of tr-TDTS data}

\noindent\textbf{A. Transient phase separation}

\begin{figure}[h]
    \includegraphics[]{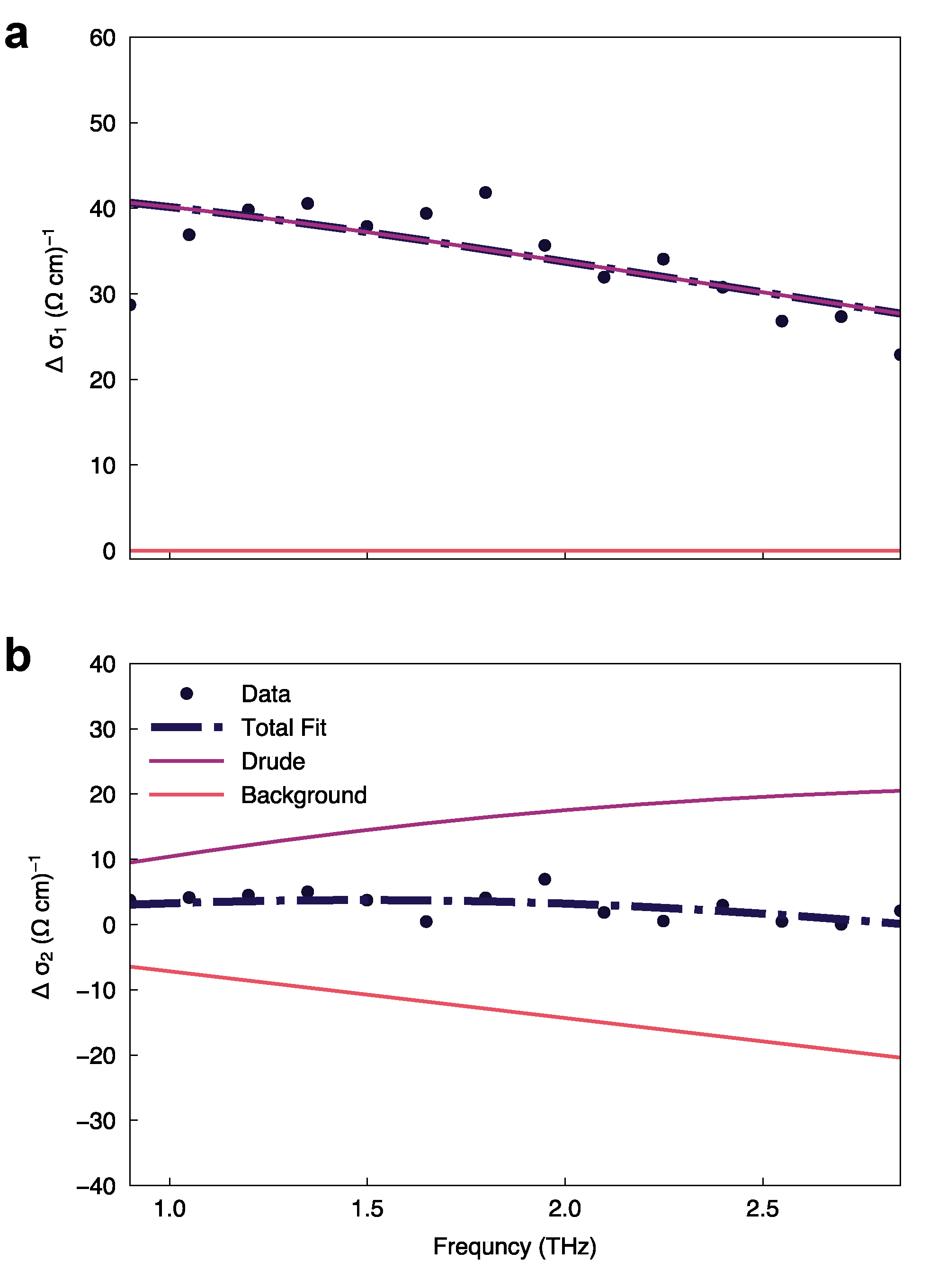}
    \label{Drude_only}
    \caption{\textbf{Drude-Lorentz Fitting of the transient changes to the optical conductivity at $t$ = 0.3 ps.} \textbf{a},\textbf{b} Photo-induced changes to the real (a) and imaginary (b) parts of the optical conductivity. The solid lines represent various parts of the Drude-Lorentz fit (Methods), while the dot-dashed line indicates the total fit.}
    
\end{figure}

In this section, we show that our photo-induced optical spectra cannot be explained by the formation of a phase-separated state following the photo-doping process. To do this, we rely on the effective medium approximation to simulate the response of a heterogeneous photo-excited medium. We start by solving the Bruggeman formula to determine a model for the effective dielectric response of the phase-separated material. Since we are considering a mixture of metallic and insulating phases, we included two terms giving \cite{stroud_effective_1998}:

\begin{equation}
p_{1} \frac{\varepsilon_{1} - \varepsilon_{eff}}{\varepsilon_{1} + (d-1)\varepsilon_{eff}} + p_{2} \frac{\varepsilon_{2} - \varepsilon_{eff}}{\varepsilon_{2} + (d-1)\varepsilon_{eff}} = 0
\end{equation} 

\noindent where $p_{1}$ is the metallic area fraction, $p_{2}$  is the insulating area fraction, $d$ is the dimension, $\varepsilon_{1}$ is the dielectric function of the metallic areas, $\varepsilon_{2}$ is the dielectric function of the insulating areas, and $\varepsilon_{eff}$ is the effective response of the total area. Solving for $\varepsilon_{eff}$ gives:

\begin{equation}
\varepsilon_{eff} = \frac{1}{2(d-1)}\left[ d\varepsilon_{avg} -\varepsilon_{1} -\varepsilon_{2}\pm\sqrt{(d\varepsilon_{avg} -\varepsilon_{1} -\varepsilon_{2})^{2}+4(d-1)\varepsilon_{1} \varepsilon_{2}}  \right]
\end{equation} 

\noindent where $\varepsilon_{avg} = p_{1}\varepsilon_{1}+p_{2}\varepsilon_{2}$ and $p_{i}$ sum to unity. From this expression, we can retrieve the effective light-induced changes to the conductivity of the heterogeneous film: 

\begin{equation}
\Delta\sigma_{eff}=-i(\varepsilon_{eff}-\varepsilon_{2})\omega\varepsilon_{0}
\end{equation}
\noindent where $\omega$ is frequency.

To define $\varepsilon_{1}$, we note that the system only contains metallic puddles after it is photo-excited. Indeed, the dataset at the early time delay of $t=0.3$ ps can only be fit using a Drude term and a broad background term that primarily affects the imaginary part of the optical conductivity and is likely caused by higher energy changes to the conductivity (Figure S2). This result implies that the photo-induced layer is completely metallic at this time delay, suggesting that $p_{1}=1$ and  $p_{2}=0$. Accordingly, we can define the response of the metallic portions using this fitted Drude response at this early time delay (see Methods in the main text):

\begin{equation}
\Delta\sigma_{Drude} =\frac{D}{2}\left[\frac{1}{\gamma-i\omega}\right]
\end{equation}

\noindent where the fitting constants $D$ and $\gamma$ and  are the Drude strength and width, respectively. At $t$ = 0.3 ps, these values are $330\pm7$ $\Omega^{-1}\text{cm}^{-1}\text{ps}^{-1}$ and 3.9 $\pm$ 0.1 THz respectively. Thus, we get the following result for $\varepsilon_{1}$:

\begin{equation}
\varepsilon_{1}=\varepsilon_{2}+\frac{i\Delta\sigma_{Drude}}{\omega\varepsilon_{0}}
\end{equation}

\noindent $\varepsilon_{2}$ can be defined using the measured equilibrium response of the material:

\begin{equation}
\varepsilon{_2}=(n+ik)^{2}
\end{equation}
where $n$ and $k$ are the real and imaginary parts of the index of refraction, respectively (Figure 1c). However, our results do not change significantly with or without this condition. 

\begin{figure}[h]
    \includegraphics[]{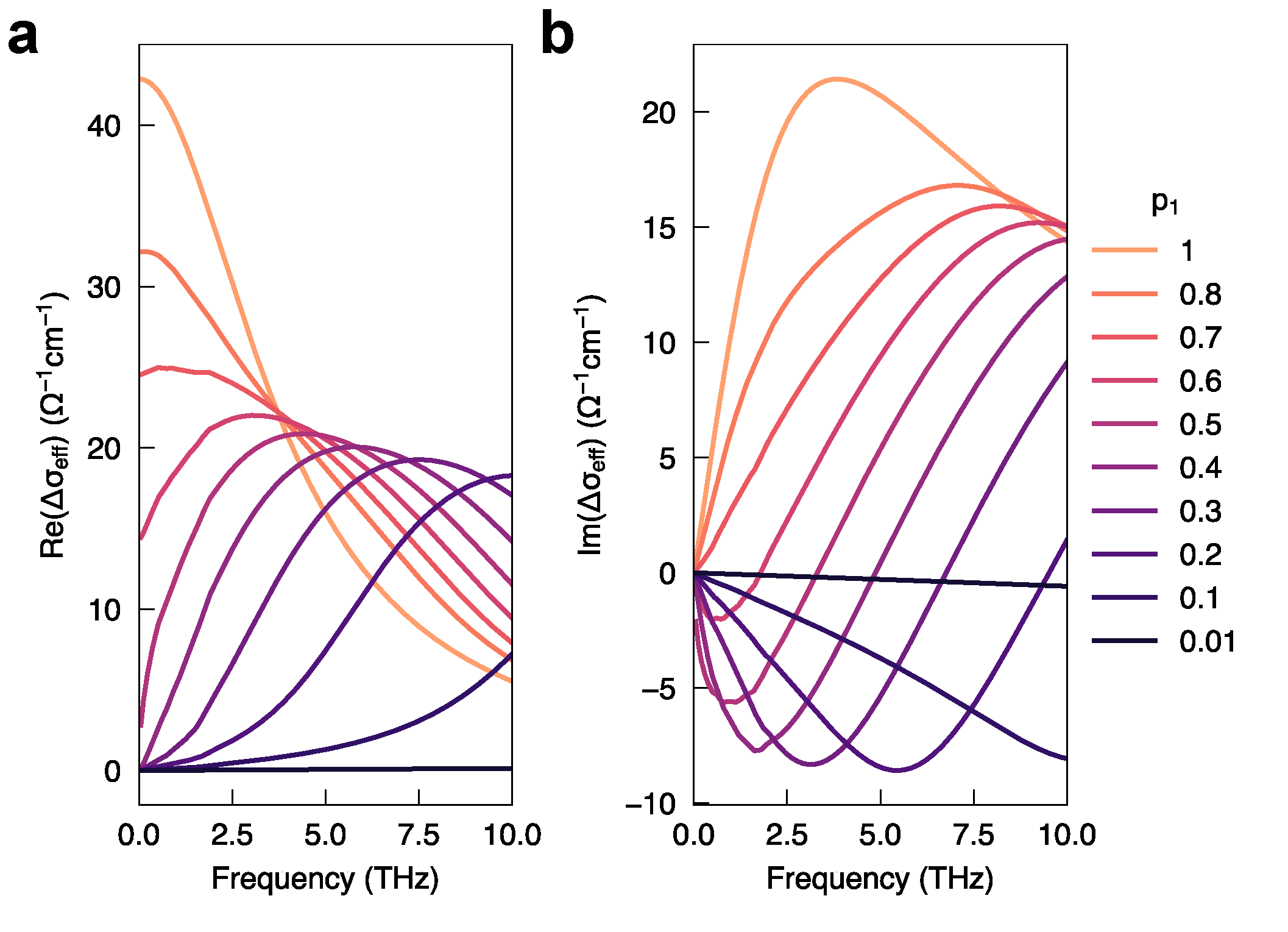}
    \label{Bruggeman1}
    \caption{\textbf{Results of the Bruggeman analysis.} \textbf{a},\textbf{b} Simulated effective light-induced changes to the real (a) and imaginary (b) parts of the optical conductivity of the heterogeneous film. These curves were generated by using experimental inputs into the Bruggeman formula and varying the volume fraction of the metallic and insulating portions of the sample.}
\end{figure}

With these considerations in mind, we can obtain a series of curves showing the expected response for the phase-separated material with different volume fractions of metallic and insulating puddles (Figure S3). Here, we enforce charge conservation by keeping $\Delta\sigma_{Drude}(\omega=0)\cdot p_{1}$ to be constant. There are several important features that contrast starkly with our measured data on both a qualitative and quantitative level, ultimately ruling out phase-separation and heterogeneity as a possible explanation of our results.

\begin{itemize}
\item{\textbf{Spectral discrepancies:} While a finite energy peak does emerge (Figure S3) as the volume fraction of the insulating phase grows, its characteristics are qualitatively and quantitatively different from the experimentally observed finite energy peak. First, the peak that emerges from the effective medium analysis is significantly broader, spanning a frequency window of larger than 10 THz. On the other hand, the width of our observed resonance is always less than 2 THz (Figure 2). Moreover, the central frequencies of the finite energy peak that emerge from the Bruggeman analysis range from right above 0 THz to beyond 10 THz, as it is extremely sensitive to the volume fraction. In contrast, the central frequency of the measured mode is always between 1.5 and 2 THz.}

\item{\textbf{Temporal dependence:} In our experimental data, we observe that the Drude spectral weight at early time delays evolves into a finite energy peak. For this to be explained by the heterogeneity hypothesis, the area fraction of the metallic regions $p_{1}$ would have to decrease as time evolves. However, as shown in Figure S3, as $p_{1}$ is decreased the finite energy peak will strongly blue-shift. It is clear from Figure 2 that our experimentally measured finite energy peak does not blue-shift, thereby ruling out this scenario. }

\begin{figure}[h]
    \includegraphics[]{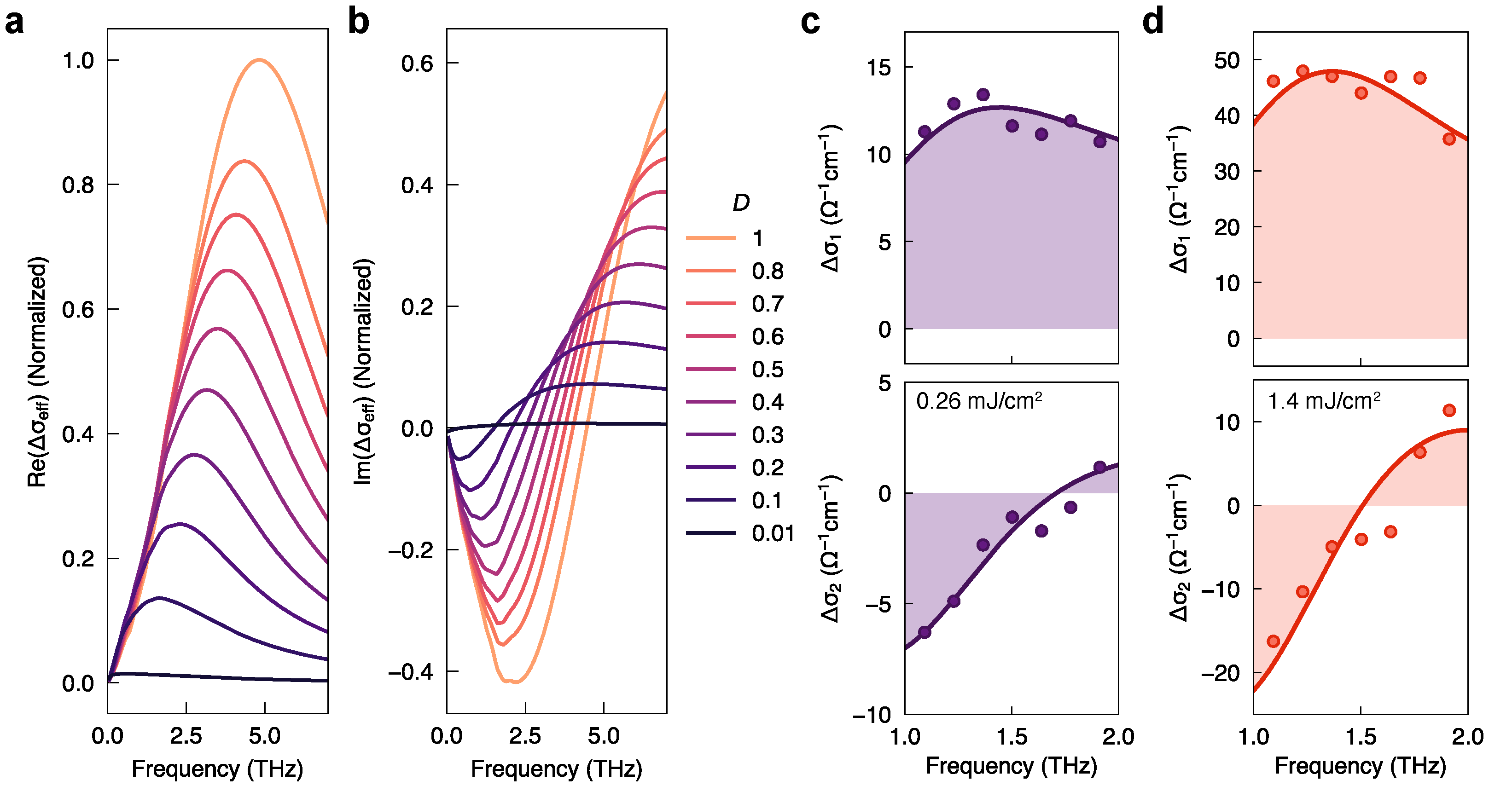}
    \label{FDEP}
    \caption{\textbf{Dependence of the experimental and simulated finite energy peaks on photo-carrier density.} \textbf{a},\textbf{b} Simulated effective light-induced changes to the real (a) and imaginary (b) parts of the optical conductivity of the heterogeneous film extracted from the Bruggeman formalism. These curves were generated by using experimental inputs into the Bruggeman formula and varying the value of the Drude strength ($D$). The values shown in the legend are normalized to the maximum $D$ used, and the curves in the plots are normalized to the $D$ = 1 curve. The value of $p_{1}$ was fixed to a representative value of 0.2. \textbf{c},\textbf{d} Real (top) and imaginary (bottom) parts of the experimental optical conductivity obtained with pump fluences of 0.26 mJ/cm$^{2}$ (c) and 1.4 mJ/cm$^{2}$ (d). This data was taken at $t$ = 2.55 ps and at 80 K.}
\end{figure}

\item{\textbf{Fluence dependence:} As mentioned above, the central frequency of the finite energy peak in the Bruggeman analysis is highly dependent on the ratio of metallic and insulating areas. We also found that it is highly sensitive to the Drude spectral weight within the metallic regions as well (Figures S4a,b). This behavior possibly stems from the fact that this feature has been attributed to the formation of a plasmon \cite{stroud_effective_1998, zeng_numerical_1989}, and is accordingly highly sensitive to the carrier density. Experimentally, we can control the carrier density by varying the pumping fluence. Thus, if the phase separation hypothesis were to correctly explain our observations, the central frequency of our observed resonance should vary drastically with the fluence of the driving pulse, strongly blue-shifting as the fluence is increased. To test this hypothesis, we performed spectrally-resolved experiments as a function of pump fluence. As seen in Figures S4c,d, we do not observe such behavior. Instead, we observe that the central frequency of the measured peak remains largely unchanged as the fluence is changed by more than 5 times, in contrast to the result predicted by the Bruggeman analysis. This result again rules out the inhomogeneity hypothesis.} 

\end{itemize}

For these reasons, we believe that our data cannot be explained by the heterogeneity hypothesis and that the Drude-Lorentz model used in our manuscript is the most accurate representation of our data. Moreover, these results are consistent with our expectations based on physical reasoning. The metallic patches in chemically doped Sr$_2$IrO$_4$ samples nucleate from the sites of the chemical dopants \cite{okada_imaging_2013}, which tend to not be distributed evenly across the sample. The insulating patches appear in regions that are away from any chemical dopants. This sort of phase separation should not appear in our experiment, which is photo-doped as opposed to chemically doped. Unlike chemical dopants, we expect that photo-dopants will initially be distributed evenly across the sample because the pump beam is homogeneous, creating a uniform response. This assumption is supported by the fact that the system is well-described by the Drude model at early time delays.\\

\noindent\textbf{B. Defect capture and charge trapping}

In the following, we provide arguments and data to rule out defect capture as a possible explanation of our observed dynamical and spectral responses.

\begin{figure}[h]
    \includegraphics[]{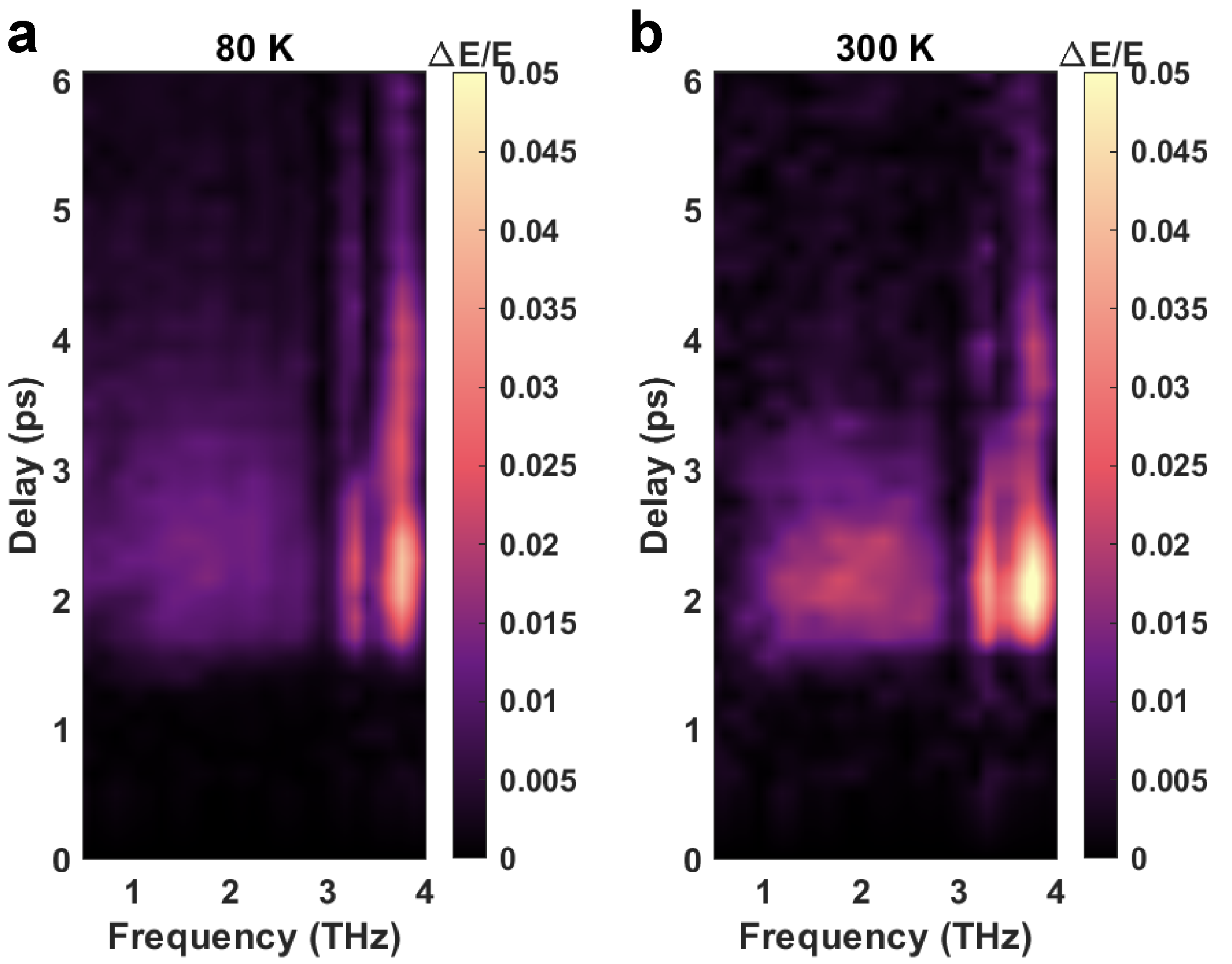}
    \label{TDEP}
    \caption{\textbf{Temperature dependence of the photo-induced changes to the optical spectrum.}  \textbf{a},\textbf{b} Transient differential optical spectra obtained at temperatures of 80 K (a) and 300 K (b). Data were taken with a pump fluence of 2 mJ/cm$^{2}$.}
\end{figure}

\noindent\textbf{\textit{Dynamics}}\\
Our picosecond-scale recombination dynamics cannot be mediated by defect capture. In this scenario, self-trapped excitons, either of intrinsic (through polaron formation) or extrinsic (through trapping by defects) origin, should show different intra-excitonic transition energies compared to free excitons due to the local trapping potential. Assuming that such behavior is at play, there are two options: (1) excitons are trapped within our accessible time and energy windows but the frequency shift is small; or (2) the peak frequency of trapped excitons is so different that it lies outside of our bandwidth. 

To rule out case (1), we first note that because the trapped excitons would have a different intra-excitonic transition energy than the free excitons, we should observe two distinct Lorentzian peaks within our bandwidth. However, as can be seen in Extended Data Figure 1, our data is well-fit by a single finite energy Lorentzian. We can further rule out scenario (1) by considering the fact that free excitons and trapped excitons typically have very different recombination times (usually much slower in the case of trapped excitons \cite{li_self-trapped_2020}). If there were contributions from both free and trapped excitons within our bandwidth, our decay dynamics would show at least two exponential components. In contrast, our decay dynamics remain well described by a single-exponential throughout our temperature, energy, and fluence range (Figure 2, Figure 3, Extended Data Figure 2). Thus, by considering both the spectral characteristics of our data as well as the temporal dynamics, we believe that we can safely rule out scenario (1). 

For case (2), the spectral weight decay observed within our bandwidth would correspond to a spectral weight transfer to a separate peak corresponding to trapped excitons that lies outside our bandwidth. The percentage of excitons that is transferred into the trapped state should decrease upon heating because thermal energy in general works against exciton trapping. Moreover, the excitons should stay free for longer times for the same reason, extending the lifetime of our observed peak. However, this contradicts our observation that the decay becomes faster at higher temperatures (Figure 3b), and that the strength of the excitonic peak does not decrease at higher temperatures (Figure S5). Both of these experimental observations contradict the expected behavior for the trapping hypothesis, thereby ruling out scenario (2). 

\noindent\textbf{\textit{Spectral response}}\\
Charge trapping is not a possible explanation for the existence of the 1.5 THz peak in the photo-induced change to the optical conductivity. If our data were to be explained by charge trapping into an impurity level, it would have to be extremely shallow, as we observe the peak frequency to be around 1.5 THz. This frequency corresponds to a temperature of 72 K. As such, we would expect that these impurity states should thermally dissociate in our experimental temperature range of 80 K to 300 K. However, we observe that the exciton remains stable across this entire temperature range, ruling out this scenario (Figure S5).\\

\noindent\textbf{C. Two-band Drude Model}

A two-band Drude model with different scattering rates cannot reproduce our observed spectra. A Drude model will always peak at zero frequency. This is true even if multiple Drude peaks are included. In contrast, in our data we found that at most time delays the spectrum reaches a maximum at the central frequency of the finite energy peak. This behavior suggests that the spectrum is described by the superposition of a Drude peak and a finite energy peak. Moreover, at early time delays (e.g. 0.3 ps) when little Lorentzian spectral weight has formed, a single Drude term is enough to fit our dataset (Figure S2), further ruling out the hypothesis of multiple Drude peaks. As such, we do not believe that a multi-band Drude model could explain the existence of the finite energy peak, nor could it explain the ultrafast spectral weight transfer from the Drude component into the Lorentzian component. We note that our fits are highly constrained, as we simultaneously fit both the real and imaginary parts of the optical conductivity. \\

\noindent\textbf{D. Strong bosonic coupling}

For the following reasons, the 1.5 THz peak observed in our transient optical conductivity spectra cannot be attributed to a sideband arising from strong coupling to bosonic modes. First, the peak should be separated from the Drude peak by around the bosonic mode energy \cite{basov_electrodynamics_2011}. However, as seen in Figure 1c, there are no optically active bosonic
modes in our measured energy region. Although sideband formation does not necessarily require the boson to be optically active, we note that previous inelastic x-ray scattering experiments \cite{dashwood_momentum-resolved_2019, sen_absence_2022} on Sr$_2$IrO$_4$ measured no optical phonon modes at all (optically active or inactive) in the energy range of interest over the full Brillouin zone. While there are acoustic modes spanning the energy range of 0 to 5 meV \cite{dashwood_momentum-resolved_2019, sen_absence_2022}, their density of states is vanishingly small \cite{dashwood_momentum-resolved_2019, parlinski_lattice_2005}. Moreover, the formation of a sideband arises from the coupling of electrons to a bosonic mode with well-defined energy, often modeled as a dispersionless Einstein phonon mode exhibiting a large phonon density of states at one particular energy. In contrast, a highly dispersive acoustic phonon (or magnon) mode has a featureless density of states, which is incompatible with our observation of a Lorentzian peak with well-defined energy in the optical conductivity spectrum.

Second, the dynamics of the sideband should be similar to the dynamics of the main Drude peak (with a possible time delay in formation) as the spectral weight of the sideband is being redistributed from the Drude spectral weight \cite{basov_electrodynamics_2011}. It is clear from Figure 2 that the Drude and Lorentzian components of our data show completely different dynamics. Crucially, at later time delays ($t > 1$ ps), the Drude spectral weight is vanishing while the Lorentzian spectral weight is still near its peak (Figure 2). Moreover, between 0.5 and 1 ps, the two show opposite trends, with the Drude spectral weight decreasing and the Lorentzian spectral weight increasing. We believe that these dynamics rule out the possibility of electron-boson coupling as an explanation of the observed 1.5 THz mode. 

Third, if the observed Lorentzian peak were a sideband of a Drude peak, this would imply a Drude peak width below our lowest measured frequency of 0.35 THz (Fig. S1). However, chemically-doped samples show a Drude peak that is much broader than 0.35 THz \cite{xu_optical_2020}, and our THz data at short time delays show that photo-doped samples exhibit a Drude peak that is much broader than 0.35 THz (See Fig. 2b and Fig. S2). The temporal evolution of our THz data also shows that the Drude peak does not suddenly sharpen to below 0.35 THz after the onset of the Lorentzian component. We are not aware of any physical mechanism that would cause such a drastic change of the photo-induced Drude response. Such a scenario would be rather exotic, as Drude peaks sharper than 0.35 THz (1.45 meV) are usually only resolved in the rare cases of heavy fermion systems \cite{Scheffler2005} and nearly-perfect or highly-mobile conductors \cite{armitage_electrodynamics_2018}. Neither of these scenarios apply to Sr$_2$IrO$_4$.

Finally, electron-boson coupling should generate a unique spectral response that can be detected through an extended Drude model analysis of the optical conductivity data. The effective mass $m^{*}(\omega)$ should be enhanced below the bosonic mode frequency, while the scattering rate $1/\tau(\omega)$ should begin to increase above the bosonic mode frequency. These quantities can be experimentally extracted using the following relations \cite{armitage_electrodynamics_2018}: 

\begin{equation}
\begin{split}
m^*(\omega) &\propto -\frac{1}{\omega}\text{Im}\left[\frac{1}{\tilde{\sigma}(\omega)}\right]\\
\frac{1}{\tau(\omega)} &\propto \text{Re}\left[\frac{1}{\tilde{\sigma}(\omega)}\right]\\
\end{split}
\end{equation}

We calculated the frequency dependent effective mass and scattering rate using Equation S7 for a characteristic dataset showing the Lorentzian peak. The results are shown in Figure S6. Within our resolution $1/\tau(\omega)$ is not enhanced above some characteristic energy and remains rather constant. More importantly, $m^*$ is mostly below 0, indicating a breakdown of the model. These results further rule out the electron-boson coupling scenario.

\begin{figure}[h]
    \includegraphics[width=.75\textwidth]{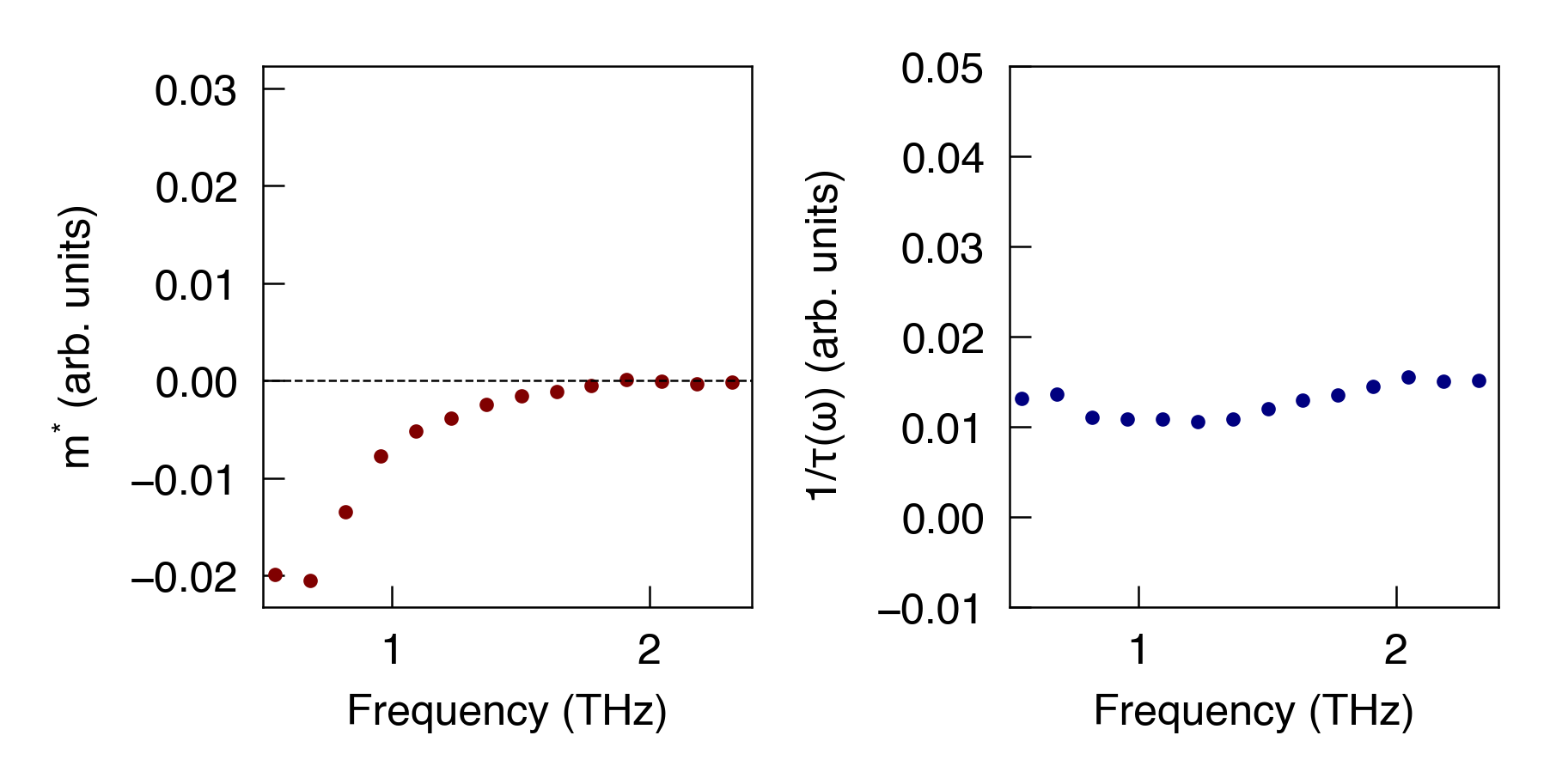}
    \label{extended_drude}
    \caption{\textbf{Results of the extended Drude model analysis.} $m^*(\omega)$ (left) and $1/\tau(\omega)$ (right) calculated using Equation S7. This data was collected with an $\alpha$-resonant pump at a fluence of 2.0 mJ/cm$^2$ and $t$ = 1.2 ps. The temperature was 80 K.}
\end{figure}

\noindent\textbf{E. Pseudogap formation}

In chemically doped Sr$_2$IrO$_4$ samples, the characteristics of the pseudogap are highly dependent on both the number of charge carriers that are doped into the system as well as the temperature of the sample \cite{kim_fermi_2014, kim_observation_2016, de_la_torre_collapse_2015}. Thus, to determine whether a photo-induced pseudogap might explain our data, we can compare the temperature and pump fluence dependence of our photo-induced spectra with the spectra of chemically doped Sr$_{2}$IrO$_{4}$. 

We find that both the temperature and fluence dependencies are not consistent with the expected behavior of the pseudogap. At a fixed temperature below the onset of the pseudogap, previous works found that the energy scale of the pseudogap can vary drastically with the doping level, reaching up to 80 meV \cite{kim_fermi_2014}. On the other hand, as shown in Figure S4, the energy scale of our observed finite energy peak does not change as the fluence - and thus the photo-dopant density - is varied. 

Similarly, the pseudogap energy scale is highly dependent on the temperature of the sample, forming below a critical temperature around 110 K for a reported electron concentration of $\sim$3 $\%$ \cite{kim_fermi_2014} and enhancing as the temperature is reduced \cite{kim_fermi_2014, kim_observation_2016}. On the other hand, in our case we observe that the energy scale of the finite energy peak remains unchanged as the temperature is varied (Figure S5). This observation is at odds with the physics of pseudogap formation. 

Moreover, we find that the spectral characteristics of our finite energy peak are not compatible with pseudogap physics. First, the energy scale of our feature is roughly 6 meV, which is much smaller than the value of the pseudogap observed by photoemission spectroscopy (50-80 meV are typical maximum values over the reported doping and temperature ranges). Second, as shown in Extended Data Figure 1, our finite energy peak is perfectly consistent with the Lorentzian lineshape, indicating that it originates from a dipole-active optical transition as opposed to a pseudogap feature. This can be confirmed by comparing against the optical response in the pseudogap phase, for example in the cuprates \cite{uchida_optical_1991} or in Sr$_{2}$IrO$_{4}$ \cite{seo_infrared_2017, xu_optical_2020}. In both cases, the low-energy response in the pseudogap phase is marked by a finite but spectrally flat optical response. 

Given these reasons, we believe that the simplest explanation for our transient optical conductivity spectra would be the formation of excitons, which is consistent not only with our spectral properties as a function of energy, temperature, and photo-dopant density, but also is the more likely scenario given the simultaneous presence of both holes and electrons. 

\newpage
\noindent\textbf{III. Determination of transient optical conductivity}

\begin{figure}[h!]
    \includegraphics[width=85mm]{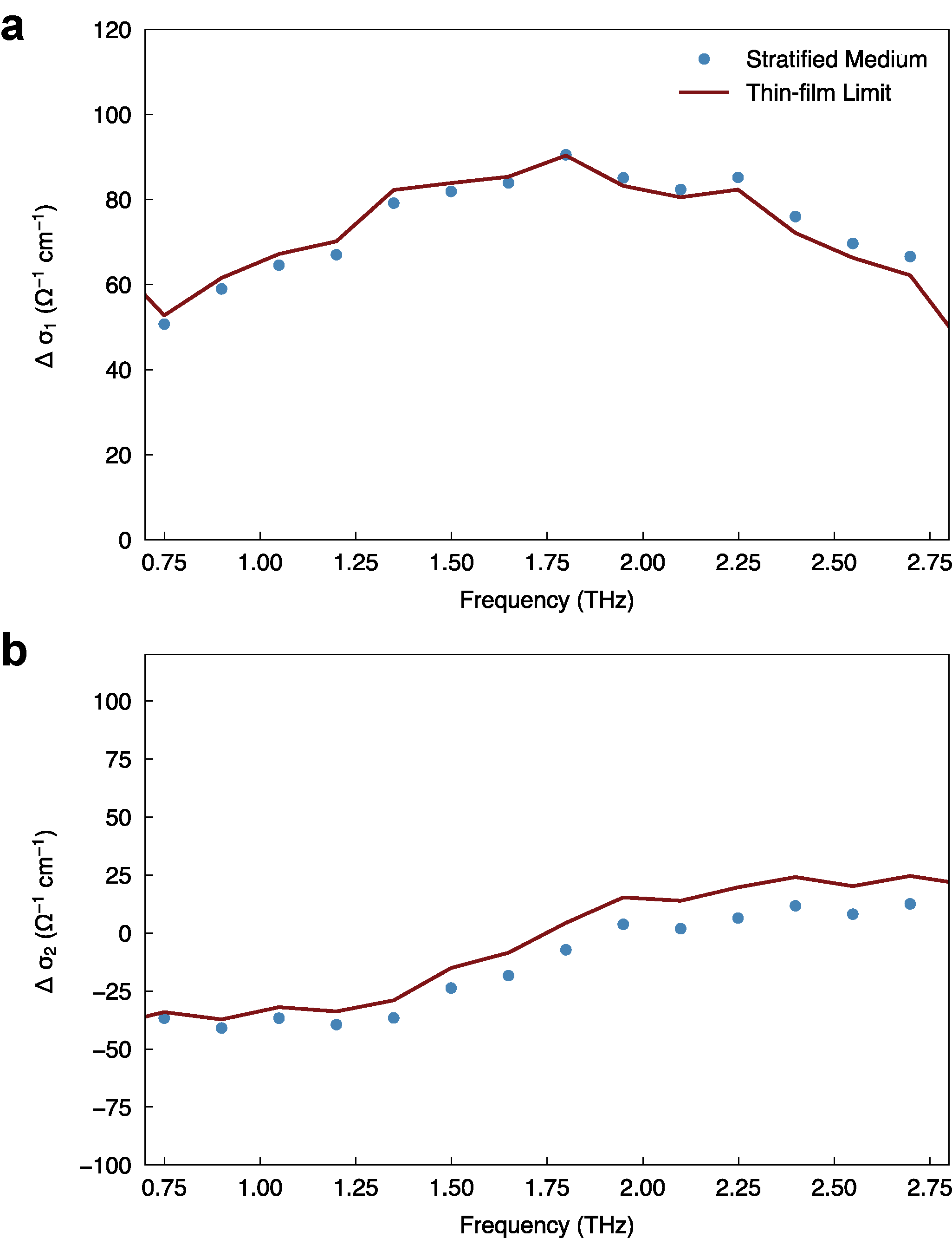}
    \label{ModelComparison}
    \caption{\textbf{Comparison of different models for extraction of transient optical conductivity.} \textbf{a},\textbf{b} Comparison of the real (a) and imaginary (b) parts of the optical conductivity obtained with the thin-film approximation and stratified medium approach. }
\end{figure}

The penetration depth mismatch between the pump and probe must be accounted for to correctly extract the transient optical conductivity. The thin-film approximation used in the manuscript (Methods) can be safely adopted due to the large penetration depth mismatch between the IR pump (73 nm) \cite{bhandari_electronic_2019} and the THz probe, which transmits through the $\sim$100 $\mu$m thick sample. To prove that the thin film approximation is accurate, we extracted the transient optical conductivity using a numerically exact, albeit more complicated, model called the stratified medium approach \cite{hunt_manipulating_2015}. This model assumes the pump decays exponentially inside the sample with a characteristic length-scale equal to the penetration depth. We break this exponential profile into 100 layers. Then, we assume that each layer has a unique photo-induced index of refraction that approaches the bulk index of refraction as the layers go deeper into the sample:

\begin{equation}
\tilde{n}^\prime\left(z,\omega\right)=\tilde{n}\left(\omega\right)+\Delta \tilde{n}\left(\omega\right)e^{-\alpha z}
\end{equation}

\noindent where $\omega$ is the frequency, $z$ is the layer depth into the sample, $\alpha$ is the penetration depth of the pump, $\tilde{n}\left(\omega\right)$ is the complex equilibrium index of refraction, and $\Delta\tilde{n}\left(\omega\right)$ is the complex photo-induced change to the index of refraction. The reflection from this stratified medium is modeled using a transfer matrix formalism, and the difference between the experimental photo-induced change in reflectivity and the modeled result is minimized using a least squares algorithm with $\Delta\tilde{n}\left(\omega\right)$ serving as the variable of interest \cite{hunt_manipulating_2015}. The results are summarized in Figure S7. The numerically solved stratified medium solution is nearly identical with the analytical thin film approximation, especially for the real part of the optical conductivity which determines the spectral weights plotted in Figure 2.

\newpage

\noindent\textbf{IV. Pump wavelength dependence}

\begin{figure}[h]
    \includegraphics[width=167mm]{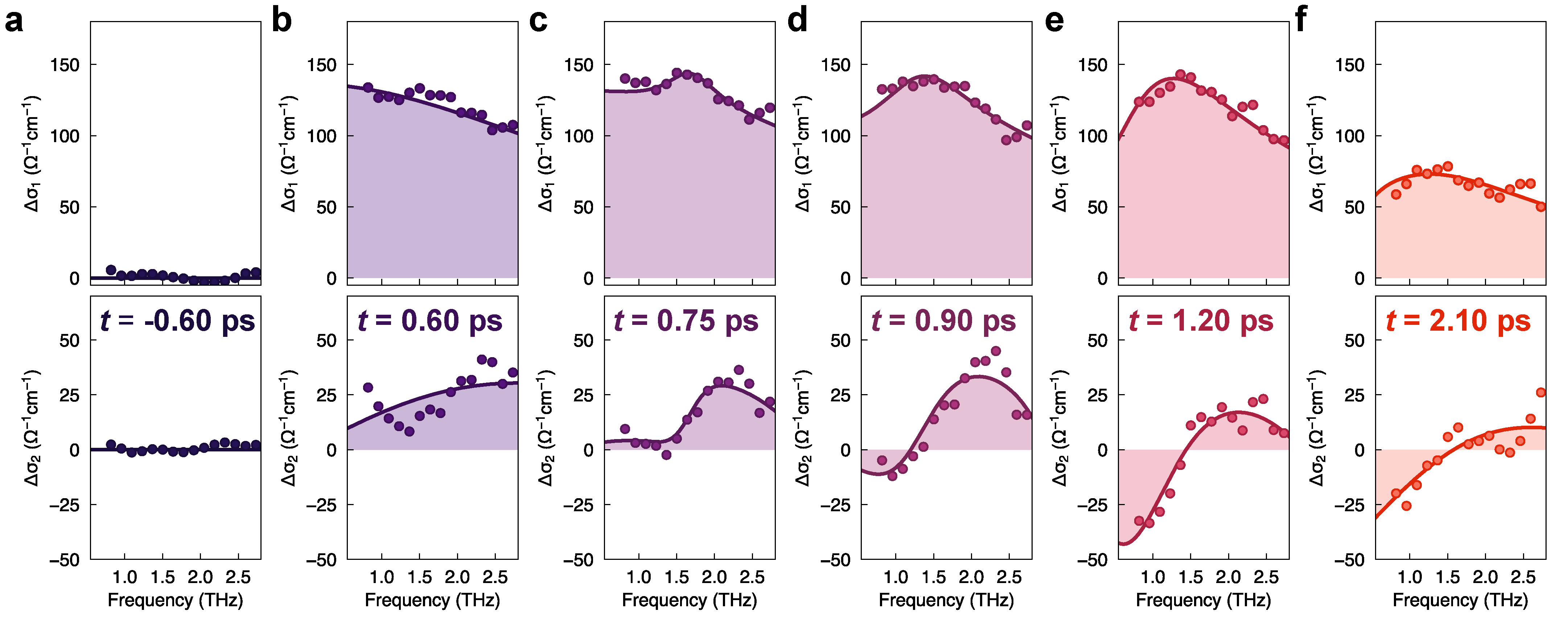}
    \label{1250_raw}
    \caption{\textbf{Photo-doping induced optical conductivity transients of Sr$_{2}$IrO$_{4}$ pumped resonant with the $\beta$ transition.} \textbf{a-f} $\Delta\sigma_{1}(\omega)$ (top panels) and $\Delta\sigma_{2}(\omega)$ (bottom panels) extracted from differential THz spectra at various $t$. The pump was set to 1.0 eV (resonant with the $\beta$ transition) and the data were taken at 80 K. Fits to the Drude-Lorentz model (Methods) are displayed as solid lines. 
}
\end{figure}
Strong spin-orbital coupling in Sr$_{2}$IrO$_{4}$ produces a filled  $J_{eff}$ = 3/2 band and a half-filled  $J_{eff}$ = 1/2 band. The latter splits into lower and upper Hubbard bands due to the on-site Coulomb interaction. In the main text, we pump the transition from the lower-to-upper Hubbard, known as the $\alpha$ transition. However, another possibility would be to pump the $J_{eff}$ =3/2 to upper Hubbard band transition, known as the $\beta$ transition. Doing so will initially excite a $J_{eff}$ = 3/2 electron to the $J_{eff}$ = 1/2 upper Hubbard band, leaving behind a hole in the $J_{eff}$ = 3/2 band. However, it is well known that in Sr$_{2}$IrO$_{4}$ the photo-carrier relaxation towards the gap edge occurs on an ultrafast ($\sim$10 fs) timescale \cite{hsieh_observation_2012}. Due to this rapid thermalization process, we should expect that some subset of the holes produced in the $J_{eff}$ = 3/2 band should relax into the lower Hubbard band, leaving them available for Hubbard exciton formation with electrons in the upper Hubbard band. However, because some subset of the $J_{eff}$ = 3/2 holes will directly recombine with $J_{eff}$ = 1/2 electrons, the number of excitons generated relative to the total number of photo-excited carriers should be reduced as compared to the case of directly pumping the $\alpha$ transition. Thus, we can compare the fraction of the total photo-carriers that become excitons when pumping the $\alpha$ versus $\beta$ transitions to provide further evidence for our interpretation of the data.

Accordingly, we performed additional experiments at a pump photon energy of 1 eV, which is resonant with the $\beta$ transition. In line with our expectations, we see the emergence of a Drude response immediately after the arrival of the pump followed by an ultrafast spectral weight transfer into a finite energy peak, similar to pumping the $\alpha$ transition (Figure S8). To characterize this spectral weight transfer, we performed the same Drude-Lorentz fitting described in the Methods section. In Figure S9, we plot the Drude and Lorentzian spectral weights for both the $\alpha$ and $\beta$ pumping cases. As can be seen, while a Drude-to-Lorentz spectral weight transfer is observed in both cases, the spectral weight ratio between the two terms is not the same for each pumping case. While the $\alpha$ case produces a Lorentz to Drude ratio of roughly 0.64 at their peaks, the $\beta$ case produces a ratio of 0.23. Since the Drude spectral weight is indicative of the number of free carriers and the Lorentzian spectral weight is indicative of the number of bound carriers \cite{kaindl_transient_2009}, we can conclude that the conversion efficiency from free carriers to bound pairs is significantly reduced in the $\beta$ case as compared to the $\alpha$ case. This is consistent with the expected result described in the previous paragraph, providing further evidence for our assignment of the Lorentzian component to an excitonic origin. 
\newpage

\begin{figure}[h]
    \includegraphics[width = \textwidth]{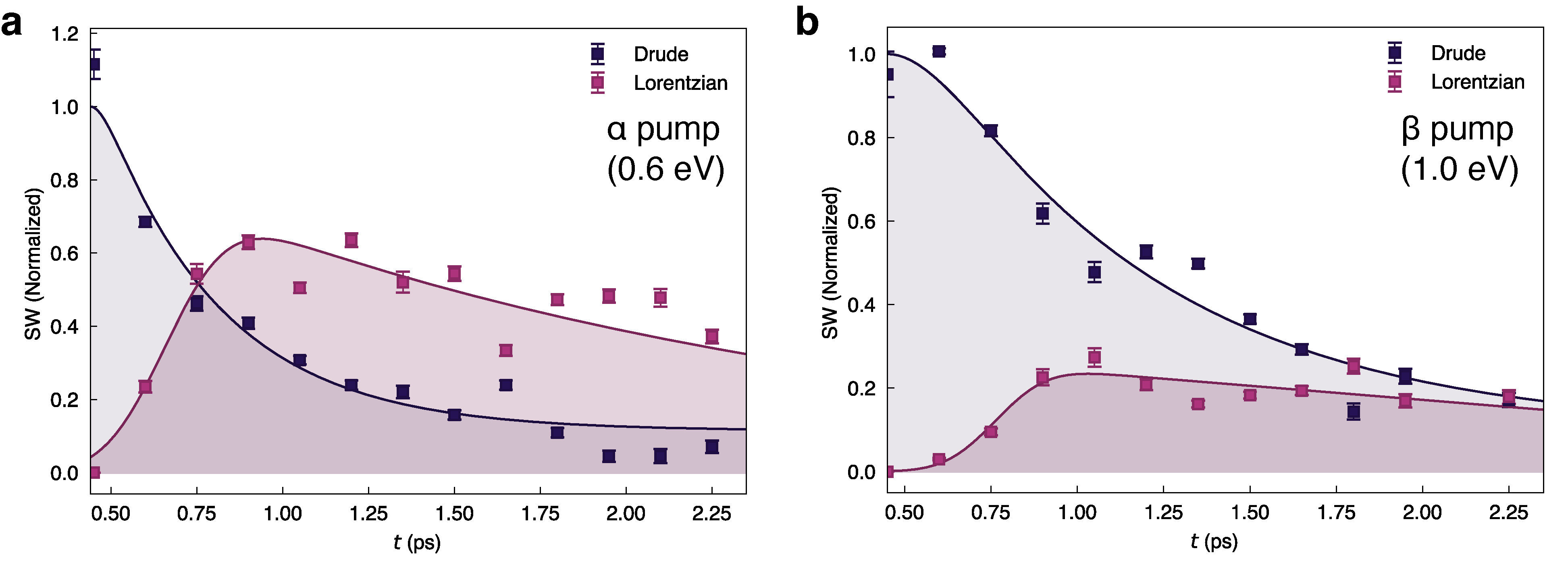}
    \label{1250_SW}
    \caption{\textbf{Temporal evolution of the spectral weight in Sr$_{2}$IrO$_{4}$.} \textbf{a},\textbf{b} Spectral weight of the Drude and HE Lorentzian terms versus time delays $t$ for the $\alpha$ pumping case (a) and the $\beta$ pumping case (b). The data is normalized to the maximum of the Drude spectral weight in each of the panels. The solid lines and shaded regions are guides to the eye. Error bars are obtained from the standard deviation of the least-squares-fitting algorithm.
}
\end{figure}

\newpage

\noindent\textbf{V. System size dependence of the microscopic model}

\begin{figure}[h]
    \includegraphics[width = \textwidth]{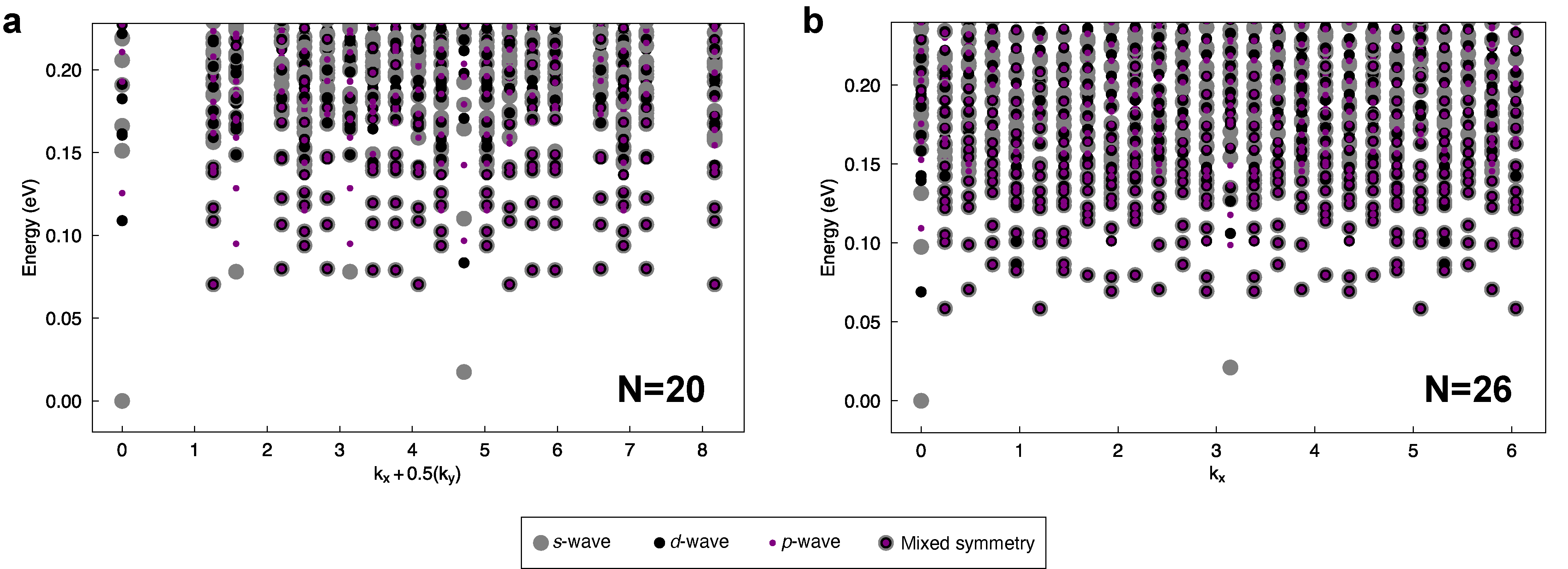}
    \label{NDEP}
    \caption{\textbf{System size dependence of the calculated spectra.} \textbf{a,b} Spectra obtained from the exact diagonalization procedure performed for two system sizes, $N = 20$ (a) and $N = 26$ (b). The latter is projected onto the $k_x$ axis. Due to the symmetry of the $N=20$ lattice, the spectrum is projected onto the $k_x + 0.5(k_y)$ axis to prevent the overlap of different states from different momenta. Large gray markers are states with $s$-wave symmetry, medium black markers are states with $d$-wave symmetry, and small violet markers are states with $p$-wave symmetry. Mixed symmetry states have multiple markers overlapped with one another. We define 0 eV to be the energy of the lowest-energy state.}
\end{figure}

In Figure 4 and Extended Data Figure 4, we present spectra obtained by exactly diagonalizing the $tJV$ model (Methods) on a $N=26$ site lattice. Here, we show the dependence of the spectra as a function of the system size $N$. In Figure S10, the results for $N=20$ and $N=26$ are shown. We find the existence of several excitonic states below a densely spaced continuum for both system sizes. These excitons have definite symmetries at high symmetry points ($\vec{k}= [0,0], [0,\pi], [\pi,0], [\pi,\pi]$ for $N=20$ and $\vec{k}=[0,0], [\pi,\pi]$ for $N=26$) and mixed symmetries at other $\vec{k}$ (Methods). However, a few details change depending on the values of $N$. Most importantly, the binding energies of the excitons, the number of excitonic states, and the order in which different symmetries of excitons appear at the high symmetry points of the spectrum can change. The latter two can be particularly sensitive to $N$, since the shapes of the lattices, which feature periodic boundary conditions, can favor different symmetries. We believe that this is a clear finite size effect that would eventually diminish with larger system sizes. 

We believe that our conclusions remain qualitatively robust despite these finite size effects, since there are always optically-allowed intra-excitonic transitions that appear at a fraction of $J$, in agreement with the peak observed in experiment, regardless of these choices. Also, we find that the excitons appear to be more stable (more clearly below the continuum) as the system size is increased.

\end{document}